\RequirePackage{fix-cm}
\documentclass[column]{svjour3}          

\smartqed  
\usepackage{mathptmx}      
\usepackage{amssymb}
\usepackage{amsmath, amscd, amsfonts}
\usepackage{graphicx}
\usepackage{latexsym}
\usepackage{float}
\usepackage[numbers]{natbib}
\usepackage{caption}

\numberwithin{equation}{section}

\usepackage{color}
\definecolor{blue1}{rgb}{0.0, 0.0, 1.0}
\definecolor{gray}{rgb}{0.9,0.9,0.9}
\definecolor{gray1}{rgb}{0.7,0.7,0.7}
\definecolor{gray2}{rgb}{0.8,0.8,0.8}
\definecolor{magenta}{rgb}{1.0, 0.0, 1.0}
\definecolor{darkred}{rgb}{0.7, 0.0, 0}

\usepackage[utf8]{inputenc}
\usepackage[english]{babel}
\usepackage{tikz}
\usetikzlibrary{shapes.geometric,fit,arrows,babel}

\usepackage{hyperref}
\hypersetup{ colorlinks=true, urlcolor  = blue, linkcolor = blue,
citecolor = blue1,}
\newcommand{\txtd}{\text{d}}

 \journalname{Frontiers in Computational Neuroscience:}

\begin{document}

\title{Optimal self-induced stochastic resonance in multiplex neural networks: electrical versus chemical synapses 
}
\subtitle{}

\author{Marius E. Yamakou \and  Poul G. Hjorth \and Erik A. Martens 
}


\institute{
M. E. Yamakou \at Max-Planck-Institut f\"{u}r Mathematik in den Naturwissenschaften, Inselstr. 22, 04103 Leipzig, Germany\\ 
M. E. Yamakou $\cdot$  P. G. Hjorth  $\cdot$ E. A. Martens \at Department of Applied Mathematics and Computer Science, Technical University of Denmark, 2800 Kgs. Lyngby, Denmark\\
E. A. Martens  \at Department of Biomedical Science, University of Copenhagen, 2200 Copenhagen, Denmark\\
\email{yamakoumarius@gmail.com}
}

\date{Received: 1 April 2020 / Accepted: 28 May 2020}

\maketitle

\begin{abstract}

Electrical and chemical synapses shape the dynamics of neural networks and their functional 
roles in information processing have been a longstanding question in neurobiology. 
In this paper, we investigate the role of synapses on the optimization of the phenomenon of 
self-induced stochastic resonance in a delayed multiplex neural network by using analytical and numerical methods. 
We consider a two-layer multiplex network, in which at the intra-layer level neurons are coupled either by electrical 
synapses or by inhibitory chemical synapses. For each isolated layer, computations indicate that weaker electrical and 
chemical synaptic couplings are better optimizers of self-induced stochastic resonance. In addition, regardless of the 
synaptic strengths, shorter electrical synaptic delays are found to be better optimizers of the phenomenon than shorter 
chemical synaptic delays, while longer chemical synaptic delays are better optimizers than longer electrical synaptic 
delays --- in both cases, the poorer optimizers are in fact worst.  It is found that electrical, inhibitory, or excitatory 
chemical multiplexing of the two layers having only electrical synapses at the intra-layer levels can each optimize the phenomenon. 
And only excitatory chemical multiplexing of the two layers having only inhibitory chemical synapses at the intra-layer levels can 
optimize the phenomenon. These results may guide experiments aimed at establishing or confirming the mechanism of self-induced 
stochastic resonance in networks of artificial neural circuits, as well as in real biological neural networks.
\end{abstract}


\section{Introduction}\label{section1}

Noise is an inherent part of neuronal dynamics and its 
effects can be observed experimentally in neuronal activity at different spatiotemporal scales, 
e.g., at the level of ion channels, neuronal membrane potentials, local field potentials, 
and electroencephalographic or magnetoencephalographic measurements~\cite{guo2018functional}.
While noise is mostly undesirable in many systems, it is now widely accepted that its presence is 
crucial to the proper functioning of neurons in terms of their information processing capabilities. 

Some mechanisms for optimal information processing are provided via the well-known and extensively studied phenomena of stochastic resonance 
(SR)~\cite{benzi1981mechanism,longtin1993stochastic,gammaitoni1998stochastic,lindner2004effects,zhang2015stochastic} and 
coherence resonance (CR)~\cite{lindner2004effects,hu1993two,pikovsky1997coherence,lindner1999analytical,neiman1997coherence,beato2007coherence,
hizanidis2008control,liu2010multiple,bing2011coherence,gu2011coherence}; or  via the lesser-known phenomenon of 
self-induced stochastic resonance  
(SISR)~\cite{freidlin2001stable,muratov2005self,deville2007nontrivial,deville2007self,yamakou2017simple,yamakou2018coherent} whose mechanism 
remains to be confirmed experimentally in real neural systems. 
Although these noise-induced phenomena may exhibit similar dynamical behaviors, each of them has different 
dynamical preconditions and emergent mechanisms, and may therefore play different functional roles in information processing. 
For further details behind the mechanisms of SR and CR, we refer the reader to references given above. We also note that 
the control of SR and CR in neural networks has attracted a lot of attention. In particular, it has been shown that 
hybrid synapses and autapses (i.e., characterized by  both electrical and chemical coupling) could be effectively used to 
control SR and CR \cite{yilmaz2013stochastic,yilmaz2016autapse}.

In this paper, we focus on self-induced stochastic resonance (SISR). SISR can occur when a multiple timescale excitable dynamical system is driven 
by vanishingly small noise. During SISR, the escape time of trajectories from one attracting region in phase space to 
another is distributed exponentially, and the associated transition frequency is governed by an activation energy. 
Suppose the system describing the neuron is placed out-of-equilibrium, and its activation energy decreases monotonically as the 
neuron relaxes slowly to 
a stable quiescent state (fixed point); then, at a specific instant during the relaxation, the timescale of escape events and 
the timescale of relaxation match, 
and the neuron fires at this point almost surely. 
If this activation brings the neuron back out-of-equilibrium, the relaxation stage can start over again, and the scenario repeats 
itself indefinitely, leading to a 
cyclic coherent spiking of the neuron which cannot occur without noise.
SISR essentially depends on 
(i) strong timescale separation between the dynamical variables; 
(ii) vanishingly small noise amplitude; 
(iii) a monotonic activation energy barrier; 
(iv) and most importantly, the periodic matching of the slow timescale of neuron's dynamics to the timescale characteristic to the noise.
Thus, compared to CR and SR, the conditions to be met for observing SISR are more subtle: Like CR, SISR does not require an external 
periodic signal as in SR. Remarkably, unlike CR, SISR does not require the neuron's parameters be close to the bifurcation thresholds,
making it more robust to parameter tuning than CR. Moreover, unlike both SR and CR, SISR 
requires a strong timescale separation between the neuron's dynamical variables.

The mechanism behind SISR suggests that in an excitable neuron, the level of noise embedded in the neuron's synaptic input 
may be decoded into a (quasi-) deterministic and coherent signal. 
To exemplify, in a network of neurons in a quiescent state (without any activity), the action of a sufficiently weak synaptic 
noise amplitude could occasionally generate a spike in each neuron. 
These spikes will have random phases, so that their total input on each individual neuron may average to a stationary random 
signal of low intensity. If the noise amplitude then suddenly increases due 
to a change in the synaptic input, the neurons may switch to the noise-assisted oscillatory mode. This can further increase 
the effective noise amplitude, so that the oscillatory mode may persist even 
after the disturbance is removed and the entire neural network in a dormant state may wake up from the outside rattle. 
 The phenomenon of SISR in neural networks could therefore play important functional roles in the regulation of the 
sleep-wake transition~\cite{patriarca2012diversity,booth2014physiologically,pereda2014electrical}.

Communication between neurons occurs through synaptic interactions. 
Two main types of synapses may be identified in neural networks, electrical synapses and chemical synapses~\cite{pereda2014electrical}. 
The corresponding functional form of the bidirectional interaction mediated by the electrical synapses is defined as the difference 
between the membrane potentials of two adjacent neurons; 
thereby making the coupling mediated by electrical synapses to be local. 
While chemical synaptic interaction always take place unidirectionally, with the signal conveyed chemically 
via neurotransmitter molecules through the synapses; thereby making chemical synaptic couplings nonlocal. 
The functional form of the chemical synaptic interaction is 
considered as a nonlinear sigmoidal input-output function~\cite{greengard2001neurobiology}. 
Moreover, chemical synapses can be inhibitory or excitatory. When an inhibitory pre-synaptic neuron spikes, 
the post-synapses neuron connected to it is prevented from spiking. When an excitatory neuron spikes, 
it induces the post-synaptic neuron to spike. In real biological neurons, the distance 
between pre- and post-synaptic ends is approximately 3.5 nm in electrical synapses, 
and comparatively large, nearly 20-40 nm~\cite{hormuzdi2004electrical} in chemical synapses. 
Distances between pre- and post-synaptic ends induce time delays in neural networks with the 
time delays of electrical synapses being generally shorter than those of chemical synapses. 

It is well known from magnetic resonance imaging that neural networks may exhibit several types of coupling schemes: 
neurons coupled via electrical synapses only;
neurons coupled via chemical synapses only; and neurons coupled by both electrical and chemical 
synapses ---  so-called hybrid 
synapes~\cite{majhi2019chimera,bera2019spike,galarreta1999network,gibson1999two,hestrin2005electrical,
galarreta2001electrical,connors2004electrical,yilmaz2013stochastic}. 
Moreover, multiplex networks of neurons can be formed from different network layers depending on their connectivity 
through a chemical link or by an ionic channel. In brain networks, different 
regions can be seen connected by functional and structural neural networks~\cite{de2017multilayer,pisarchik2014critical,andreev2018coherence}. 
In a multiplex network, each type of interaction between the nodes 
is described by a single layer network and the different layers of networks describe 
the different modes of interaction. Multilayer networks~\cite{pisarchik2014critical} open up new possibilities of optimization, 
allowing to regulate neural information processing by means of the interplay between the neurons' 
dynamics and multiplexing~\cite{battiston2017multilayer,crofts2016structure}. 
Optimization based on multiplexing could have many advantages. 
In particular, the coherent spiking activity 
of one layer (induced for example by SISR) can be optimized by adjusting the parameters 
of another layer. This is important from the point of view of engineering and brain surgery 
since it is not always possible to directly access the desired layer, while the network with which 
this layer is multiplexed may be accessible and adaptable.

Several studies have shown that multiplex networks can generate patterns with significant differences from those observed 
in single-layer networks~\cite{kouvaris2015pattern,berner2020birth,majhi2017chimera,majhi2016chimera}. 
Their use in the optimization and control of dynamical behaviors have therefore attracted much attention recently. 
The multiplexing of networks has been shown to control many dynamical 
behaviors in neural networks including 
synchronization~\cite{gambuzza2015intra,singh2015synchronization,leyva2017inter,zhang2017incoherence,andrzejak2017generalized}, pattern 
formation~\cite{kouvaris2015pattern,ghosh2016emergence,ghosh2016birth,maksimenko2016excitation,bera2017chimera,bukh2017new,ghosh2018non}, 
solitary waves~\cite{mikhaylenko2019weak} and chimera 
states~\cite{ghosh2018non,panaggio2015chimera,Scholl2016,omelchenko2018control,sawicki2019chimeras,ghosh2019taming}. 
Chimera states are synchronization patterns occurring in symmetric networks (on average), 
characterized by the coexistence of varying synchronization levels side-by-side. 
They have been shown to exist in mechanical and chemical experiments~\cite{MartensThutupalli2013,Tinsley2012,Totz2017} 
and are thought play an important role in neural systems~\cite{majhi2019chimera,bera2019spike,andrzejak2016all}. In particular, synchronization patterns 
such as chimera states occur in networks with community structure where connections are all-to-all, but coupling strengths 
are modulated so that the inter-coupling between communities (layers) are weak/sparse compared to their 
intra-coupling~\cite{Bick2019,abrams2008,MartensPanaggioAbrams2016,Martens2016} --- a configuration 
that bears strong similarity with the multilayer structure. Chimera states in such networks are of interest 
as they are multistable~\cite{Martens2010bistable} and thus configurable, so that they in principle can be 
employed to solve functional tasks such as computations~\cite{BickMartens2015} and routing of 
information~\cite{Deschle2019} in the brain. Moreover, community  networks 
of QIF neurons exhibit synchronization patterns that have been demonstrated viable  for memory 
storage and recall~\cite{Schmidt2018}. 
However, the optimization of noise-induced resonance mechanisms in neural networks based on the 
multiplexing approach have only very recently attracted attention. 
The few research works investigating the optimization of CR in neural networks are those of Semenova 
and Zakharova~\cite{semenova2018weak} and  Yamakou and Jost~\cite{yamakou2019control}. 

In~\cite{semenova2018weak} it is shown that connecting a one-layer network exhibiting CR in a multiplex way to another 
one-layer network, i.e., multiplexing, allows to control CR in the 
latter layer network. In particular, it is found that multiplexing induces CR in the network that do not demonstrate this
phenomenon in isolation. Moreover, it has been shown that CR 
can be achieved even for weak multiplexing between the layers. Surprisingly, it has also been shown that the multiplex-induced 
CR in the layer which is deterministic in isolation can 
manifest itself even more strongly than the CR in the noisy layer. However, the work in~\cite{semenova2018weak} considers 
only instantaneous synaptic connections, while it is well known that 
synaptic time delays (not negligible in neural networks) have crucial effects in neural information processing. 

Yamakou and Jost~\cite{yamakou2019control} considered synaptic time delays and 
their role in optimizing CR in a layer affected by another layer via multiplexing, which already exhibits optimal CR or SISR. 
In an isolated layer, it was shown that shorter 
synaptic time delays combined with weaker synaptic strengths optimize CR. While in the multiplex network configurations, 
stronger synaptic strengths combined with shorter synaptic time 
delays between layers induce and optimize CR in the layer where this phenomenon is non-existent in isolation. Moreover, 
their numerical simulations indicate that even at very long multiplexing 
time delays, weak (but not too weak) multiplexing strengths between the layers can induce and optimize 
CR in the layer where it is non-existent in isolation. Interestingly, 
it was further shown that with the occurrence of a different resonance phenomenon (i.e., SISR) in one layer, weak multiplexing even at 
very short synaptic time delays completely fails to optimize CR in the other layer where latter phenomenon does not exist in isolation.
This behavior further confirms the fact that even though SISR and CR lead to the the occurrence of the same 
dynamical behavior (i.e., coherent noise-induced spiking activity) in neurons in the excitable regime, they are 
fundamentally different in their dynamical and emergent nature~\cite{deville2005two}; 
in particular, SISR and CR also lead to different behaviors in multiplex networks, and possibly therefore play different functional 
roles in neural information processing.

The optimization of CR in neural networks based on the multiplexing approach have been so far studied only 
in~\cite{semenova2018weak} and~\cite{yamakou2019control}. 
A study on the optimization of SISR in neural networks based on the multiplexing approach is still lacking. 
Moreover, in~\cite{semenova2018weak} and~\cite{yamakou2019control}, 
the coupling between the neurons are mediated only by electrical synapses. 
The role of chemical synapses in the optimization of noise-induced resonance mechanisms should be equally important. 
Therefore, the aim of this paper is to study the optimization of SISR based on the multiplex approach of neural networks connected through 
time-delayed electrical and chemical synapses. In particular, we wish to address the following main questions:
\begin{enumerate}
 \item [(i)] Can SISR occurring in one layer of a multiplex network be used to optimize SISR in the another layer 
 where the phenomenon non-existent in isolation?
 \item [(ii)] What combinations of intra- and inter-layer synaptic strengths and time delays best optimize of SISR?
 \item [(iii)] Which type (electrical, inhibitory, or excitatory) of synapses is best 
 optimizer of SISR within an isolated layer and in the multiplex configuration?
\end{enumerate}

The rest of the paper is organized as follows: In Sec.~\ref{section2}, we present the mathematical 
model equations, and we explain and motivate the different configurations considered. 
In Sec.~\ref{section3}, we briefly describe the numerical methods used in simulations and analysis.
In Sec.~\ref{section4}, we consider an isolated single layer of neurons, coupled either by electrical 
synapses or chemical synapses. For both types of coupling we 
analytically establish the necessary conditions in terms of noise amplitudes and timescale separation parameter that allow 
us to observe SISR.
In Sec.~\ref{section5}, we systematically investigate synaptic parameterizations which best optimize SISR in an isolated 
layer in which the neurons are coupled either by electrical synapses or by inhibitory chemical synapses.   
We will then compare the optimization of SISR by electrical and inhibitory chemical synapses.
In Sec.~\ref{section6}, we consider multiplexed layer networks using numerical simulations. 
Having identified which synaptic configurations deteriorate SISR the most 
in isolated layers, we use the multiplexing between a first layer, where SISR is optimal and a second layer where 
SISR is non-optimal (very poor or even non-existent), with the goal of optimizing SISR  in the second layer. 
For multiplex networks, we will consider the optimization of SISR in six case scenarios: 
electrical, inhibitory, and excitatory multiplexing of two layers with electrical synaptic intra-connections; 
and electrical, inhibitory, and excitatory multiplexing of two layers with inhibitory synaptic intra-connections. 
Finally, we summarize and conclude our findings in Sec.~\ref{section7}.
\section{Mathematical model}\label{section2}

\begin{figure}[htp!]
\begin{center}
\includegraphics[width=0.4\textwidth]{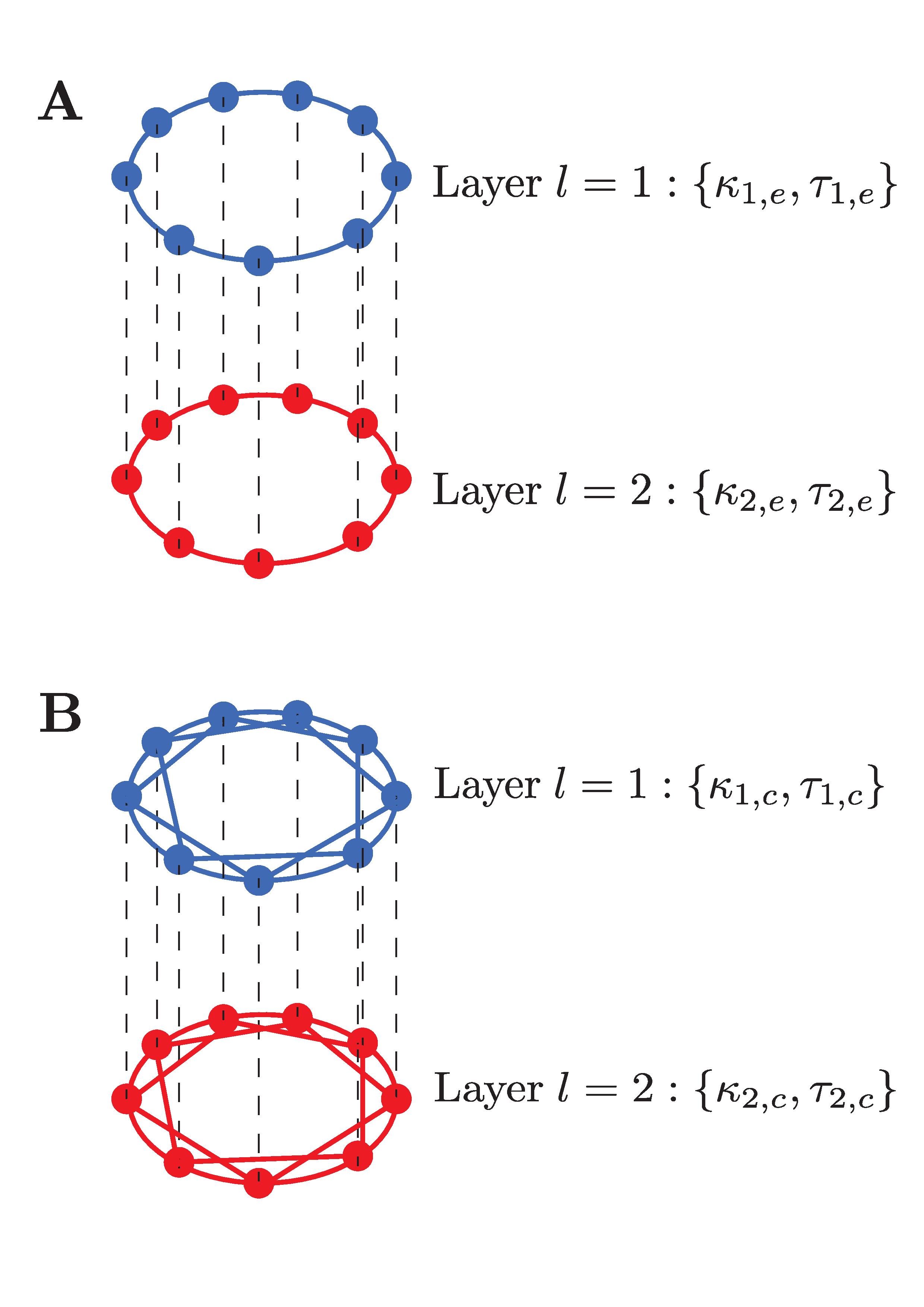}
\caption{Neurons are connected in a multiplex network with two layers $l=1,2$. 
Neurons within a layer are coupled in a ring, while neurons in adjacent layers are connected only to their adjacent neurons.   
  Scenario \textbf{(A)}: Neurons within each layer are coupled in a ring and interact only with their nearest neighbors via electrical synapses 
  ($\kappa_{_{1,e}}, \tau_{_{1,e}}, \kappa_{_{2,e}}, \tau_{_{2,e}}$); multiplexing between these layers, represented by the black vertical
  dashed lines, may occur via electrical $(\kappa_{_{m,e}},\tau_{_{m,e}})$ or (inhibitory or excitatory) chemical synapses 
  $(\kappa_{_{m,c}}, \tau_{_{m,c}})$.   
  Scenario \textbf{(B)}: Neurons within each layer are coupled in a ring and interact with $n_{l,c}$ nearest neighbors via 
  inhibitory chemical synapses $(\kappa_{_{1,c}}, \tau_{_{1,c}},\kappa_{_{2,c}},\tau_{_{2,c}})$; 
  multiplexing between these layers, represented by the black vertical dashed lines, may occur via electrical $(\kappa_{_{m,e}},\tau_{_{m,e}})$ 
  or (inhibitory or excitatory) chemical $(\kappa_{_{m,c}}, \tau_{_{m,c}})$ synapses. 
  In both scenarios, an enhanced SISR in layer $ l = 1 $ is used to optimize a poor or non-existent SISR in layer $ l = 2 $ 
  by variation of the time-delay coupling parameters within a population, 
  $(\kappa_{_{1,e}}, \tau_{_{1,e}},\kappa_{_{2,e}}, \tau_{_{2,e}}, \kappa_{_{1,c}},\tau_{_{1,c}},\kappa_{_{2,c}},\tau_{_{2,c}})$
  and between populations, $(\kappa_{_{m,e}},\tau_{_{m,e}}, \kappa_{_{m,c}}, \tau_{_{m,c}})$.
  }\label{fig:1}
\end{center}
\end{figure}

We consider a two-layer multiplex neuronal network in the excitable regime in the presence of synaptic noise, 
as illustrated in \textbf{Figure~\ref{fig:1}}.
In our study, we consider one of the simplest network topologies -- a ring network topology within layers, and a multiplex network between 
these layers such that they contain the same number of neurons and the interaction between the layers are allowed only for replica neurons. 
Each layer consists of $N$ identical FitzHugh-Nagumo (FHN) neurons~\cite{fitzhugh1961impulses,hodgkin1952quantitative}, 
connected in a ring by either only 
electrical synapses or inhibitory chemical synapses, while the inter-connections between 
layers can be either via electrical, inhibitory chemical synapses, or excitatory chemical synapses. 
It is important to point out that excitatory chemical synapses are found to induce, via time-delayed coupling bifurcations, 
a self-sustained spiking activity in the network 
of FHN neurons (each in the excitable regime) even in the complete absence of noise. We want to avoid such regimes --- 
those in which the deterministic network can oscillate due to some time-delayed coupling induced bifurcations --- 
as the coherent oscillations induced by SISR should be 
due only to the presence of noise and not because of the occurrence of bifurcations. 
For this reason, the excitatory chemical synapses are used in the optimization of SISR only when 
they do not induce oscillatory behaviors in the deterministic network, i.e., only in the multiplexing connections 
with carefully chosen synaptic strengths and time delays.

Real electrical synapses mediate bidirectional interactions and transfer signals only between neighboring neurons; 
in contrast, chemical synapses convey information unidirectionally between distantly situated neurons. 
To account for this, the model implements layers with bidirectional electrical coupling  with nearest neighbor 
interactions (\textbf{Figure~\ref{fig:1}A}), while unidirectional chemical coupling is implemented with nonlocal 
interactions, i.e., also including connections other than nearest neighbor interactions (\textbf{Figure~\ref{fig:1}B}). 
These coupling topologies and interaction modes are biologically relevant and will also allow us to compare the functional 
role played by chemical and electrical synaptic interactions in processing information generated during SISR.

The stochastic differential equations resulting from this two-layer FHN neural network are given by
\begin{align}\label{eq:1}
\begin{split}
\left\{\begin{array}{lcl}
\txtd v_{_{l,i}} &=& \left(v_{_{l,i}}-\displaystyle{\dfrac{v_{_{l,i}}^3}{3}}-w_{_{l,i}}+{E}_{_{l,i}} +{M}_{_{l,i}}^e
- {C}_{_{l,i}} - {M}_{_{l,i}}^c\right)\txtd t + \sigma_{l}\,\txtd W_{_{l,i}},\\[4.0mm]
\txtd w_{_{l,i}}&=&\varepsilon(v_{_{l,i}}+\alpha-\beta w_{_{l,i}})\,\txtd t,
\end{array}\right.
\end{split}
\end{align}
where each neuron is represented by a node $i = 1,...,N$ in the multiplex network with layers $ l = 1,2 $, and 
the functional dependencies are given by,
\begin{eqnarray}\label{eq:2}
\begin{split}
\left\{\begin{array}{lcl}
{E}_{_{l,i}}=\dfrac{\kappa_{_{l,e}}}{2n_{_{l,e}}}\sum\limits_{j=i-n_{_{l,e}}}^{i+n_{_{l,e}}}\left(v_{_{l,j}}(t-\tau_{_{l,e}})-v_{_{l,i}}(t)\right),\\[4.0mm]
{C}_{_{l,i}}=\dfrac{\kappa_{_{l,c}}}{2n_{_{l,c}}}(v_{_{l,i}}(t)-V_{_\text{syn}})\sum\limits_{j=i-n_{_{l,c}}}^{i+n_{_{l,c}}}\left\{1 + 
\exp\left[-\lambda\left(v_{_{l,j}}(t-\tau_{_{l,c}}) - \Theta_{_\text{syn}}\right)\right]\right\}^{-1},\\
{M}_{_{1,i}}^e=\kappa_{_{m,e}}\left(v_{_{2,i}}(t-\tau_{_{m,e}})-v_{_{1,i}}(t)\right),\\[2.0mm]
{M}_{_{2,i}}^e=\kappa_{_{m,e}}\left(v_{_{1,i}}(t-\tau_{_{m,e}})-v_{_{2,i}}(t)\right),\\[2.0mm]
{M}_{_{1,i}}^c=\kappa_{_{m,c}}\left(v_{_{1,i}}(t)-V_{_\text{syn}}\right)\left\{1 + \exp\left[-\lambda\left(v_{_{2,i}}(t-\tau_{_{m,c}}) - \Theta_{_\text{syn}}\right)\right]\right\}^{-1},\\
{M}_{_{2,i}}^c=\kappa_{_{m,c}}\left(v_{_{2,i}}(t)-V_{_\text{syn}}\right)\left\{1 + \exp\left[-\lambda\left(v_{_{1,i}}(t-\tau_{_{m,c}}) - \Theta_{_\text{syn}}\right)\right]\right\}^{-1}.
\end{array}\right.
\end{split}
\end{eqnarray} 
We fixed the number of neurons per layer to $ N = 25 $ throughout this study.
The membrane potential and the recovery current variables of neuron $i$ in layer $l$ are given by $v_{_{l,i}}\in\mathbb{R}$ and $w_{_{l,i}}\in\mathbb{R}$, 
respectively, and $0<\varepsilon\ll1$ sets the 
timescale separation between the fast membrane potential and the slow recovery current variables. 
The excitability threshold $\beta>0$ of the neurons is a codimension-one Hopf bifurcation parameter. $\alpha\in(0,1)$ is a constant parameter. 
The additive noise term $\txtd W_{_{l,i}}$ represents mean-centered Gaussian noise with $\langle \txtd W_{_{l,i}}(t) \, \txtd W_{_{l,i}}(t')\rangle_t = \delta(t-t')$ and variance (strength) 
$\sigma_{_l}$, and models the synaptic fluctuations observed in neural networks. 
 
${E}_{_{l,i}}$ represent the electrical synaptic interactions between neurons coupled within a ring layer network, with strength $\kappa_{_{l,e}}$ and time delay $\tau_{_{le}}$, respectively, 
and with an interaction range set to $ n_{_{l,e}}=1 $ since electrical synapses interact only locally. The coupling mediated by electrical synapses is of diffusive type, i.e., the electrical 
coupling term (intra- or inter-layer) vanishes if $v_{_{1,i}}$ and $v_{_{1,j}}$ (resp. $v_{_{2,i}}$ and $v_{_{2,j}}$) or $v_{_{1,i}}$ and $v_{_{2,i}}$ are equal. 

${M}_{_{1,i}}^e$ and ${M}_{_{2,i}}^e$ represent the coupling between layers via electrical synapses (i.e., electrical multiplexing of layers) with strength $\kappa_{_{m,e}}$ 
and delay $\tau_{_{m,e}}$, respectively.

${C}_{_{l,i}}$ represent chemical synaptic interactions between neurons coupled within a layer with ring topology, with strength $\kappa_{_{l,c}}$ and time delay $\tau_{_{l,c}}$, 
respectively, and where $1 < n_{_{l,c}} < (N-1)/2$ represents interaction range on the ring network layer;
we fix $n_{_{l,e}}=8$ all through this paper. The chemical synaptic function is modeled by a sigmoidal input-output function, 
$\Gamma(v_i)=\dfrac{1}{1 + e^{-\lambda(v_{i} - \Theta_{_\text{syn}})}}$, see Eq.~\eqref{eq:2} in Ref.~\cite{greengard2001neurobiology}, where parameter $\lambda = 10.0$ determines the slope 
of the function and $\Theta_{_\text{syn}} = - 0.25$ the synaptic firing threshold. 

${M}_{_{1,i}}^c$ and ${M}_{_{2,i}}^c$ represent the coupling between layers mediated by chemical synapses (i.e., chemical multiplexing of layers) with $\kappa_{_{m,c}}$ and $\tau_{_{m,c}}$ 
representing the strength and time delay, respectively. $V_{_\text{syn}}$ represents the synaptic reversal potential, for
$V_{_\text{syn}} < v_{_{l,i}}(t)$ the chemical synaptic interaction has a 
depolarizing effect that makes the synapse inhibitory, and for $V_{_\text{syn}} > v_{_{l,i}}(t)$ , 
the synaptic interaction has a hyper-polarizing effect making the synapse excitatory. 
For the version of the FHN neuron model used in this study, the membrane potentials 
$|v_{_{l,i}}(t)| \leq 2.0$ ($l=1,2; i = 1, 2, . . . , N$) for all time $t$. 
For the choice of fixed $V_{_\text{syn}} = -3.0$ (maintained through out our computations), 
the term ($v_{_{l,i}}(t)-V_{_\text{syn}}$) in Eq.~\eqref{eq:2} is always positive. 
So, the inhibitory and excitatory natures of chemical synapses will depend only on the sign in front of the synaptic coupling strengths 
$\kappa_{_{l,c}}$ and $\kappa_{_{m,c}}$. To make the chemical synapse inhibitory, we chose a negative sign
i.e., when the pre-synaptic neuron spikes, it prevents the post-synaptic neuron from spiking and conversely, 
a positive sign for excitatory chemical synapses.

\section{Numerical methods}\label{section3}

In our numerical simulations, we used the fourth-order Runge-Kutta algorithm for stochastic processes~\cite{kasdin1995runge} to integrate  over a very long time interval 
($T=600'000$ time units) to average time series over time with $7$ realizations for each noise amplitude. In the numerical simulations, this long time interval permitted us to 
collect with a small noise amplitude at least $125$ interspike intervals with $\varepsilon=0.0005\ll1$. Each network layer had $N=25$ neurons.

To measure how pronounced SISR is,  we used the coefficient of variation ($R_{_{T}}$), which is an important statistical measure based on the time intervals between spikes.
It measures the regularity of noise induced spiking and therefore a measure of how pronounced SISR can be at a 
particular noise amplitude. $R_{_{T}}$ exploits the inter-spike interval (ISI) where the $m$th interval is defined as the 
difference between two consecutive spike times $t^m_i$ and $t^{m+1}_i$ of neuron $i$ in a network, namely $\text{ISI}_i=t^{m+1}_i-t^m_i>0$. 
For the $i$th neuron, the ratio between the standard deviation and the mean defines the coefficient of variation of the ISIs 
over a time interval $[0,T]$ as~\cite{pikovsky1997coherence}:
\begin{equation}\label{eq:9}
R_{_{T_i}}=\dfrac{\sqrt{\langle ISI_i^2\rangle-\langle ISI_i\rangle^2}}{\langle
ISI_i\rangle},
\end{equation}
where $\langle ISI_i\rangle$ and $\langle ISI_i^2\rangle$ represent
the mean and the mean squared inter-spike intervals of the $i$th neuron, respectively.
The above definition of $R_{_{T}}$ is limited to characterizing SISR in an isolated neuron.
For a network of coupled neurons, SISR can be measured by redefining $R_{_{T}}$ as follows~\cite{masoliver2017coherence}:
\begin{equation}\label{eq:11}
R_{_{T}}=\dfrac{\sqrt{\langle\overline{ISI^2}\rangle-\langle\overline{ISI}\rangle^2}}{\langle\overline{ISI}\rangle},
\end{equation}
with
\begin{eqnarray}\label{eq:10}
\begin{split}
\left\{\begin{array}{lcl}
\langle\overline{ISI}\rangle=\displaystyle{\dfrac{1}{N}}\sum\limits_{i=1}^{N}\langle ISI_i \rangle,\\[5.0mm]
\langle\overline{ISI^2}\rangle=\displaystyle{\dfrac{1}{N}}\sum\limits_{i=1}^{N}\langle ISI_i^2 \rangle,
\end{array}\right.
\end{split}
\end{eqnarray}
where the extra bar indicates the additional average over the total number of neurons $N$ in the layer.

Of course, other statistical measures exist such as the correlation time, the power spectral density, and the signal-to-noise ratio which are commonly used measures to 
quantify the coherence of noise induced spiking activity. However, from a neurobiological point of view, $R_{_{T}}$ is more important than the other measures because it is 
related to the timing precision of the information processing in neural systems~\cite{pei1996noise}. Because of $R_{_{T}}$'s importance in neural information processing, 
we shall use it to characterize the regularity of the noise-induced oscillations generated by SISR in our neural network. For a Poissonian spike train (rare and incoherent spiking), 
$R_{_{T}}=1$. If $R_{_{T}}<1$, the sequence becomes more coherent, and $R_{_{T}}$ vanishes for a periodic deterministic spike train. $R_{_{T}}$ values greater than $1$ correspond to a 
point process that is more variable than a Poisson process~\cite{yamakou2018coherent,kurrer1995noise}.

\section{Conditions for SISR in isolated layers in the excitable regime}\label{section4}
We first consider the case of isolated layers of the multiplex networks in \textbf{Figure~\ref{fig:1}A} and \textbf{B}. Thus, neurons in such an isolated layer are 
connected either only via electrical synapses or via chemical synapses. In particular, here we will establish the analytic conditions necessary for the emergence of the 
SISR in these isolated network layers of FHN neurons in the excitable regime. From these conditions, we will furthermore obtain the minimum and maximum noise amplitudes 
required for SISR to occur in an isolated layer. 

For SISR to occur it is necessary to be in the excitable parameter regime. The isolated FHN neuron has a unique and stable fixed point in this regime. Choosing an initial 
condition in the basin of attraction of this fixed point will result in at most one large non-monotonic excursion into the phase space after which the trajectory asymptotically 
approaches the fixed point and stays there until initial conditions are changed again~\cite{yamakou2018coherent,izhikevich2000neural}.

Considering the multiplex networks in \textbf{Figure~\ref{fig:1}} with disconnected layers
$(\kappa_{_{m,e}} = \kappa_{_{m,c}}=0)$, we may place an isolated neuron ($\kappa_{_{l,e}}=0$ or $\kappa_{_{l,c}} = 0$) into 
an excitable regime by fixing parameter $\alpha=0.5$. The bifurcation parameter $\beta$ is chosen such 
that $\beta>\beta_{_h}(\varepsilon)$, where $\beta_{_h}(\varepsilon)$ is defined as the Hopf bifurcation 
value of an isolated neuron. Fixing the timescale separation parameter value to $\varepsilon=0.0005$, we calculate the 
Hopf bifurcation value to be $\beta_{_h}(\varepsilon)=0.7497$. It is important to note 
that for $\beta\leq\beta_{_h}(\varepsilon)$, an isolated neuron is in the oscillatory regime --- a regime that we want to avoid since the coherent oscillations generated by SISR are due only to 
the presence of noise rather than to the occurrence of a Hopf bifurcation~\cite{yamakou2018coherent}. 

Moreover, we have to ensure that the network of coupled neurons as a whole stays in the excitable regime, rather than just 
single neurons in isolation. Indeed, certain time-delayed couplings may induce self-sustained oscillations in a  network layer 
even though the isolated neurons remain inside the excitable regime. In layers with excitatory chemical synapses, 
a saddle-node bifurcation onto a limit cycle may generate self-sustained oscillations induced via time-delayed couplings~\cite{scholl2009time}. 
On the other hand, when used for the multiplexing of layers, 
some values of time delays and coupling strengths of the excitatory chemical synapses 
cannot provoke this saddle-node bifurcation. Therefore, we did not consider excitatory 
chemical synapses for the coupling of neurons within layers, but rather only for the coupling between layers. 
Thus, we need to make sure that neurons connected in each network layer stay outside the parameter 
regime where oscillations are induced by time-delayed coupling. First, we need to determine if such a regime exists and to identify it.

\begin{figure}[htp!]
    \begin{center}
    \includegraphics[width=1.0\textwidth]{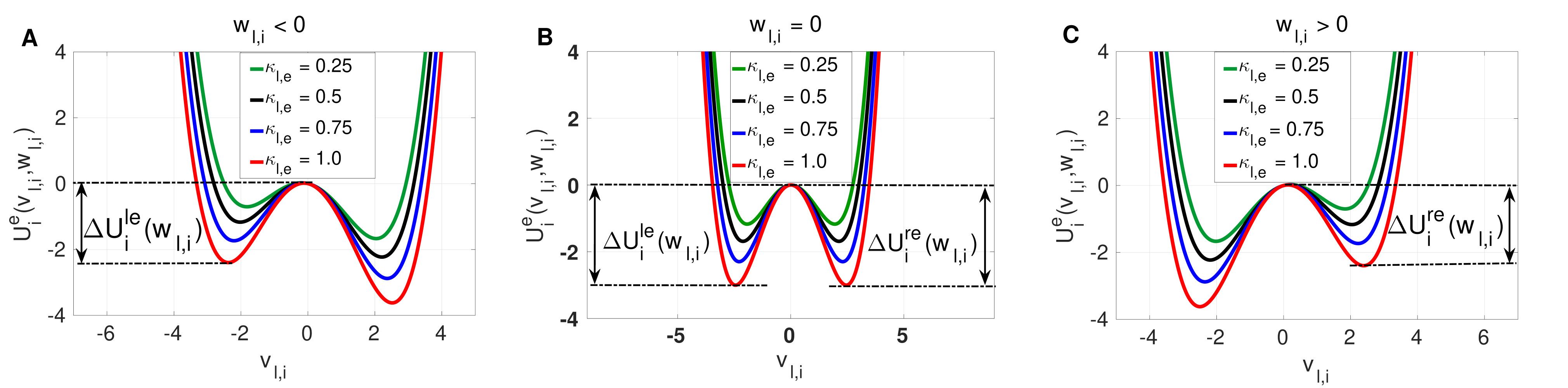}
    \caption{The electrical interaction potential $U_i^e(v_{_{l,i}},w_{_{l,i}})$ in Eq.~\eqref{eq:4a} is shown for a locally  coupled ring network topology 
    ($n_{_{l,e}}=1$) with the energy barriers for the asymmetric cases ($w_{l,i}\neq0$) (panels \textbf{(A)} and \textbf{(C)}) and symmetric ($w_{l,i}=0$) case (panel \textbf{(B)}). 
    The stronger the intra-layer synaptic strength $\kappa_{_{l,e}}$ is, the deeper the 
    energy barrier functions $\triangle U_i^{le}(w_{l,i})$ and  $\triangle U_i^{re}(w_{_{l,i}})$ are. 
    The saddle point and the left and right minima of the interaction potential are located at
    $v_{_{l,i}}=v^*_m(w_{_{l,i}})$, $v_{_{l,i}}=v^*_l(w_{_{l,i}})$, and $v_{_{l,i}}=v^*_r(w_{_{l,i}})$, respectively.
    \label{fig2}}
    \end{center}
\end{figure}

\begin{figure}[htp!]
\begin{center}
\includegraphics[width=1.0\textwidth]{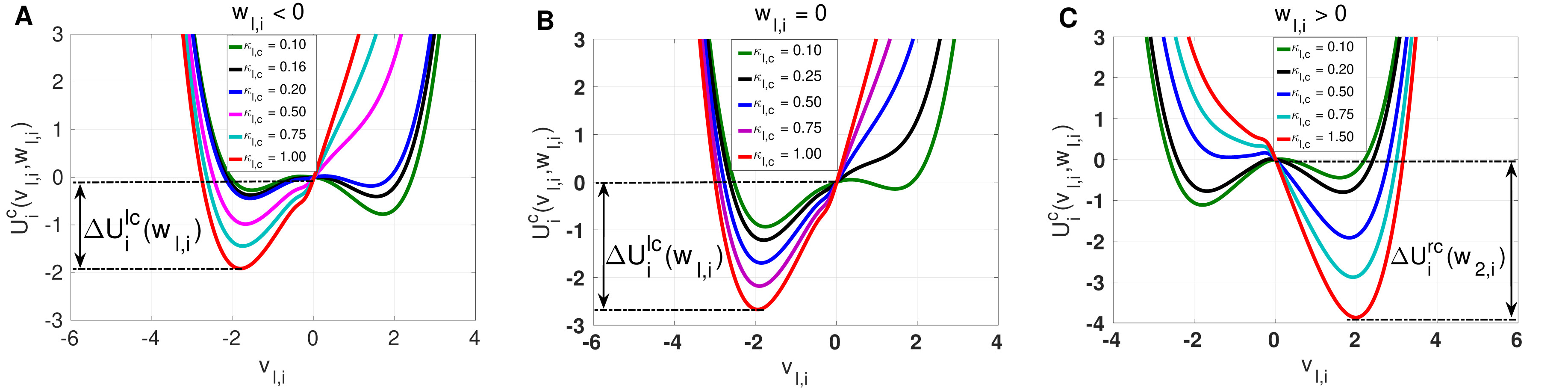}
\caption{ Landscapes of the inhibitory chemical interaction potential $U_i^c(v_{_{l,i}},w_{_{l,i}})$ in Eq.~\eqref{eq:4a} for a non-locally ($n_{l,c}=8$) 
coupled ring network topology. The symmetry of the potential is governed not by the slow variable $w_{_{l,i}}$ as in the case of the electrical interaction potential, 
but by the chemical synaptic  strength $\kappa_{_{l,c}}$. As with electrical synaptic strength, the stronger the intra-layer chemical synaptic strength $\kappa_{c}$  is, 
the deeper the energy barrier functions $\triangle U_i^{lc}(w_{_{l,i}})$ and $\triangle U_i^{rc}(w_{_{l,i}})$ are.
\label{fig3}}
\end{center}
\end{figure}

Taking the limit $\varepsilon\rightarrow0$ in the isolated layer $l=1,2$  ($\kappa_{_{m,e}} = \kappa_{_{m,c}} = 0$) for either electrical ($\kappa_{_{l,c}}=0$) or 
chemical synapses ($\kappa_{_{l,e}}=0$) only, the equations for each neuron in this layer reduces to coupled Langevin equations of the form,
\begin{eqnarray}\label{eq:3}
\displaystyle{ \txtd v_{_{l,i}}}&=&\displaystyle{ -\dfrac{\partial U^{e,c}_i(v_{_{l,i}},w_{_{l,i}})}{\partial v_{_{l,i}}} \txtd t+\sigma_l \;   \txtd  W_{l,i}},
\end{eqnarray}
where the electrical $U^e_i(v_{_{l,i}},w_{_{l,i}})$ and chemical $U^c_i(v_{_{l,i}},w_{_{l,i}})$ interaction potentials ($i=1,...,N$) are double-well potentials given by 
Eq.~\eqref{eq:4a}  and may be viewed  as functions of $v_{_{l,i}}$ where $w_{_{l,i}}$ is nearly constant.  Fig.~\ref{fig2} and  Fig.~\ref{fig3} respectively show the modulation 
of landscapes of electrical and chemical interaction potentials with changing synaptic strength. 
\begin{eqnarray}\label{eq:4a}
\begin{split}
\left\{\begin{array}{lcl}
U^e_i(v_{_{l,i}},w_{_{l,i}})&=&\dfrac{1}{12}v_{_{l,i}}^4-\dfrac{1}{2}v_{_{l,i}}^2+v_{_{l,i}}w_{_{l,i}}-
\dfrac{\kappa_{_{l,e}}}{2n_{_{l,e}}}\sum\limits_{j=i-n_{_{l,e}}}^{i+n_{_{l,e}}}\Big(v_{_{l,i}}(t)v_{_{l,j}}(t-\tau_{_{l,e}})-\dfrac{1}{2}v_{_{l,i}}(t)^2\Big),\\[5.0mm]
U^c_i(v_{_{l,i}},w_{_{l,i}})&=&\dfrac{1}{12}v_{_{l,i}}^4-\dfrac{1}{2}v_{_{l,i}}^2+v_{_{l,i}}w_{_{l,i}}\\
&+&\dfrac{\kappa_{_{l,c}}}{2n_{_{l,c}}}\sum\limits_{j=i-n_{_{l,c}}}^{i+n_{_{l,c}}}
\dfrac{1}{2}v_{_{l,i}}(t)\big(v_{_{l,i}}(t)-2V_{_\text{syn}}\big)\left\{1 + \exp\Big[-\lambda\big(v_{_{l,j}}(t-\tau_{_{l,c}}) - \Theta_{_\text{syn}}\big)\Big]\right\}^{-1},
\end{array}\right.
\end{split}
\end{eqnarray}

We observe three different behaviors for the electrical potential interaction $U^e_i(v_{_{l,i}},w_{_{l,i}})$.
(i) When $w_{l,i}<0$ we find that $U^e_i(v_{_{l,i}},w_{_{l,i}})$  is asymmetric with the shallower well on the left. 
The neuron is close to the stable homogeneous fixed point at $(v^*_{_{l,i}},w^*_{_{l,i}})=(-1.003975,-0.666651)$ and 
a spike consists of jumping over the left energy barrier $\triangle U^{le}(w_{_{l,i}})$ into the right well, see \textbf{Figure~\ref{fig2}A}. 
(ii) When $w_{_{l,i}}=0$, then $U^e_i(v_{_{l,i}},w_{_{l,i}})$ is symmetric with $\triangle U^{le}(w_{_{l,i}})=\triangle U^{re}(w_{_{l,i}})$, and 
the neuron is half way between the quiescent state and the spike state, see \textbf{Figure~\ref{fig2}B}. 
(iii) When $w_{_{l,i}}>0$,  then $U^e_i(v_{_{l,i}},w_{_{l,i}})$ is also asymmetric. 
The neuron has spiked and a return to the quiescent state (the homogeneous fixed point) consists of 
jumping over the right energy barrier $\triangle U^{re}(w_{_{l,i}})$ into the left well, see \textbf{Figure~\ref{fig2}C}. 
The intra-layer electrical synapse $\kappa_{_{l,e}}$ does not change the symmetry (or asymmetry) of the interaction potential 
$U^e_i(v_{_{l,i}},w_{_{l,i}})$. It only changes the depth of the energy barriers. The stronger $\kappa_{_{l,e}}$ is, 
the deeper the energy barrier functions $\triangle U^{le}(w_{_{l,i}})$ and  $\triangle U^{re}(w_{_{l,i}})$ defined in Eq.~\eqref{eq:5} are. 

The chemical potential interaction $U^c_i(v_{_{l,i}},w_{_{l,i}})$ shows a richer landscape dynamics due to its stronger nonlinearity.  
We first notice that just like the intra-layer electrical synaptic strength 
$\kappa_{_{l,e}}$, intra-layer inhibitory chemical synaptic strength $\kappa_{_{l,c}}$ changes the depth of the energy 
barriers $\triangle U_i^{lc}(w_{l,i})$ and  $\triangle U_i^{rc}(w_{l,i})$. That is, the stronger $\kappa_{_{l,c}}$ is, the deeper 
the energy barriers $\triangle U_i^{lc}(w_{l,i})$ and 
$\triangle U_i^{rc}(w_{l,i})$ are. In contrast to the electrical synaptic strength $\kappa_{_{l,e}}$, the inhibitory chemical synaptic 
strength $\kappa_{_{l,c}}$ is capable  of changing the symmetry or (asymmetry) of the chemical potential $U^c_i(v_{_{l,i}},w_{_{l,i}})$, where we distinguish the following cases: 
(i) When $w_{l,i}<0$, then $U^c_i(v_{_{l,i}},w_{_{l,i}})$ can be symmetric or asymmetric depending on the value of the inhibitory chemical synaptic strength $\kappa_{_{l,c}}$. 
If $w_{l,i}<0$ and $\kappa_{_{l,c}}=0.16$, we see from \textbf{Figure~\ref{fig3}A}  that $U^c_i(v_{_{l,i}},w_{_{l,i}})$ is symmetric and becomes asymmetric as $\kappa_{_{l,c}}$ changes. 
(ii) When $w_{l,i}=0.0$, we do not have any symmetric chemical potential landscape as shown in \textbf{Figure~\ref{fig3}B}, contrasting our observations for the electrical potential. 
(iii) For $w_{l,i}>0$ (see \textbf{Figure~\ref{fig3}C}),  the chemical potential landscape is symmetric for $\kappa_{_{l,c}}=0.2$ 
and the becomes asymmetric as $\kappa_{_{l,c}}$ changes.
Moreover, we notice that for values of the chemical synaptic strength $\kappa_{l,c}$ for which the chemical 
interaction potential is symmetric, the energy barriers 
functions are shallower than in the symmetric case of the electrical potential.
The important common feature of the electrical and inhibitory chemical potential is the deepening of the energy
barriers $\triangle U_i^{le,c}(w_{l,i})$ and 
$\triangle U_i^{re,c}(w_{l,i})$ with increase in the intra-layer electrical $\kappa_{_{l,e}}$ and inhibitory 
chemical $\kappa_{_{l,c}}$ synaptic strengths shall explain why SISR is deteriorated by stronger intra-layer synaptic connections.

We choose parameters of the coupled neurons in Eq.~\eqref{eq:3} such that they satisfy the conditions necessary for the occurrence of SISR. 
These conditions are adapted from those valid for an isolated FHN neuron  (\cite{yamakou2018coherent,deville2005two}) 
so that they include (one at a time)  the time-delayed electrical and inhibitory chemical synaptic connections between the FHN neurons coupled in a ring network. The resulting conditions are :

\begin{eqnarray}\label{eq:4}
\begin{split}
\left\{\begin{array}{lcl}
\displaystyle{ \lim \limits_{(\varepsilon,\sigma_l)\rightarrow (0,0)}\dfrac{\sigma_l^2}{2}\ln(\varepsilon^{-1})\in 
\Big(\triangle U_i^{le}(w^*_{_{l,i}}),F_e(\kappa_{_{l,e}},\tau_{_{l,e}},n_{_{l,e}})\Big)},\\[4.0mm]
\displaystyle{ \lim \limits_{(\varepsilon,\sigma_l)\rightarrow (0,0)}\dfrac{\sigma_l^2}{2}\ln(\varepsilon^{-1})\in 
\Big(\triangle U_i^{lc}(w^*_{_{l,i}}),F_c(\kappa_{_{l,c}},\tau_{_{l,c}},n_{_{l,c}})\Big)},\\[4.0mm]
\displaystyle{ \lim \limits_{(\varepsilon,\sigma_l)\rightarrow (0,0)}\dfrac{\sigma_l^2}{2}\ln(\varepsilon^{-1})=\mathcal{O}(1)},\\[5.0mm]
\beta-\beta_{h}(\varepsilon)>0,
\end{array}\right.
\end{split}
\end{eqnarray}
where
\begin{eqnarray}\label{eq:5}
\begin{split}
\left\{\begin{array}{lcl}
F_e(\kappa_{_{l,e}},\tau_{_{l,e}},n_{_{l,e}}):=\Big\{(\kappa_{_{l,e}},\tau_{_{l,e}},n_{_{l,e}}):\triangle U_i^{le}(w_{_{l,i}})=\triangle U_i^{re}(w_{_{l,i}})\Big\},\\[2.0mm]
F_c(\kappa_{_{l,c}},\tau_{_{l,c}},n_{_{l,c}}):=\Big\{(\kappa_{_{l,c}},\tau_{_{l,c}},n_{_{l,c}}):\triangle U_i^{lc}(w_{_{l,i}})\:\:\text{or}\:\:\triangle U_i^{rc}(w_{_{l,i}})\:\:\text{is maximum}\Big\},\\[3.0mm]
\triangle U_i^{le,c}(w_{_{l,i}}):=U_i^{e,c}\big(v^*_m(w_{l,i}),w_{_{_{l,i}}}\big)-U_i^{e,c}\big(v^*_l(w_{_{l,i}}),w_{_{l,i}}\big),\\[2.0mm]
\triangle U_i^{re,c}(w_{_{l,i}}):=U_i^{e,c}\big(v^*_m(w_{l,i}),w_{_{_{l,i}}}\big)-U_i^{e,c}\big(v^*_r(w_{_{l,i}}),w_{_{l,i}}\big),
\end{array}\right.
\end{split}
\end{eqnarray}
with
\begin{equation}\label{eq:6}
\begin{split}
v^*_{l,m,r}(w_{_{l,i}}):=\Big\{v_{_{l,i}}: v_{_{l,i}}-\displaystyle{\dfrac{v_{_{l,i}}^3}{3}}-w_{_{l,i}}+\dfrac{\kappa_{_{l,e}}}{2n_{_{l,e}}}\sum\limits_{j=i-n_{_{l,e}}}^{i+n_{_{l,e}}}\Big(v_{_{l,j}}(t-\tau_{_{l,e}})-v_{_{l,i}}(t)\Big)=0\Big\},
\end{split}
\end{equation}
for electrical synapses and 
\begin{eqnarray}\label{eq:7}
\begin{split}
v^*_{l,m,r}(w_{_{l,i}})&:=\Big\{v_{_{l,i}}: v_{_{l,i}}-\displaystyle{\dfrac{v_{_{l,i}}^3}{3}}-w_{_{l,i}}\\
&-\dfrac{\kappa_{_{l,c}}}{2n_{_{l,c}}}\sum\limits_{j=i-n_{_{l,c}}}^{i+n_{_{l,c}}}(v_{_{l,i}}-V_{_\text{syn}})\left\{1 + \exp\Big[-\lambda\Big(v_{_{l,j}}(t-\tau_{_{m,c}}) - \Theta_{_\text{syn}}\Big)\Big]\right\}^{-1}=0\Big\},
\end{split}
\end{eqnarray}
for chemical synapses. Furthermore, the solution sets of Eqs.~\eqref{eq:6} and \eqref{eq:7} are such that $v^*_l(w_{_{l,i}})<v^*_m(w_{_{l,i}})<v^*_r(w_{_{l,i}})$ define the left 
stable, middle unstable and right stable branches of the cubic nullcline of each FHN neuron.

The energy barrier functions $\triangle U_i^{le,c}(w_{_{l,i}})$ and $\triangle U_i^{re,c}(w_{_{l,i}})$ can be obtained from the electrical interaction 
potential $U^e_i(v_{_{l,i}},w_{_{l,i}})$ and the inhibitory chemical interaction potential $U^c_i(v_{_{l,i}},w_{_{l,i}})$ by taking the 
difference between the potential function value at the saddle point $v^*_m(w_{_{l,i}})$ and at the local 
minima $v^*_{l,r}(w_{_{l,i}})$ of these interaction potentials~\cite{yamakou2018coherent}. 
The energy barriers $\triangle U_i^{le}\big(w^*_{_{l,i}}\big)$ or $\triangle U_i^{lc}\big(w^*_{_{l,i}}\big)$ 
(which has to be crossed to induce a spike)  is the value of the left energy barrier 
function at the $w_{_{l,i}}$-coordinate of the  stable homogeneous steady 
state $\big[v^*_{_{l,i}}(\kappa_{_{l,e}},\tau_{_{l,e}},n_{_{l,e}}),w^*_{_{l,i}}(\kappa_{_{l,e}},\tau_{_{l,e}},n_{_{l,e}})\big]$  
or  $\big[v^*_{_{l,i}}(\kappa_{_{l,c}},\tau_{_{l,c}},n_{_{l,c}}),w^*_{_{l,i}}(\kappa_{_{l,c}},\tau_{_{l,c}},n_{_{l,c}})\big]$, respectively. 
This is where the electrical $\triangle U_i^{le}\big(w^*_{_{l,i}}(\kappa_{_{l,e}},\tau_{_{l,e}},n_{_{l,e}})\big)$ and 
chemical $\triangle U_i^{lc}\big(w^*_{_{l,i}}(\kappa_{_{l,c}},\tau_{_{l,c}},n_{_{l,c}})\big)$ energy barrier functions 
get their $\kappa_{_{l,e}}$, $\tau_{_{l,e}}$, $n_{_{l,e}}$ and $\kappa_{_{l,c}}$, $\tau_{_{l,c}}$, $n_{_{l,c}}$ dependence from. 

Now from the first two conditions of Eq.~\eqref{eq:4}, we obtain the noise amplitude range  $[\sigma^{\text{min}}_{l},\sigma^{\text{max}}_{l}]$  within which SISR occurs 
in the layer network of electrically (chemically) coupled FHN neurons: 
\begin{eqnarray}\label{eq:8}
\begin{split}
\left\{\begin{array}{lcl}
\displaystyle{\sigma^{\text{min}^{e}}_{l}=\sqrt{\dfrac{2\triangle U_i^{le}\big(w^*_{_{l,i}}(\kappa_{_{l,e}},\tau_{_{l,e}},n_{_{l,e}})\big)}{\ln(\varepsilon^{-1})}}},\\
\displaystyle{\sigma^{\text{max}^{e}}_{l}=\sqrt{\dfrac{2F_{e}(\kappa_{_{l,e}},\tau_{_{l,e}},n_{_{l,e}})}{\ln(\varepsilon^{-1})}}}.\\[5.0mm]
\displaystyle{\sigma^{\text{min}^{c}}_{l}=\sqrt{\dfrac{2\triangle U_i^{lc}\big(w^*_{_{l,i}}(\kappa_{_{l,c}},\tau_{_{l,c}},n_{_{l,c}})\big)}{\ln(\varepsilon^{-1})}}},\\
\displaystyle{\sigma^{\text{max}^{c}}_{l}=\sqrt{\dfrac{2F_{c}\big(\kappa_{_{l,c}},\tau_{_{l,c}},n_{_{l,c}}\big)}{\ln(\varepsilon^{-1})}}}.
\end{array}\right.
\end{split}
\end{eqnarray}
We observe that $\sigma^{\text{min}^e}_l$ and $\sigma^{\text{min}^c}_l$ depend on the fixed point coordinate $w^*_{l,i}(\kappa_{_{l,e}},\tau_{_{l,e}},n_{_{l,e}})$
and $w^*_{l,i}(\kappa_{_{l,c}},\tau_{_{l,c}},n_{_{l,c}})$ which in turn also depends on the synaptic parameters $\kappa_{_{l,e}},\tau_{_{l,e}},n_{_{l,e}}$ and 
$\kappa_{_{l,c}},\tau_{_{l,c}},n_{_{l,c}}$, respectively.
Therefore, changing $(\kappa_{_{l,e}},\tau_{_{l,e}}, n_{_{l,e}})$ or $(\kappa_{_{l,c}},\tau_{_{l,c}}, n_{_{l,c}})$ will 
change the value of $w^*_{l,i}(\kappa_{_{l,e}},\tau_{_{l,e}},n_{_{l,e}})$ 
or $w^*_{l,i}(\kappa_{_{l,c}},\tau_{_{l,c}},n_{_{l,c}})$, which will in turn change the value of $\sigma^{\text{min}^e}_l$ or 
$\sigma^{\text{min}^c}_l$ via the energy barrier function $\triangle U_i^{le}(w^*_{l,i})$ or $\triangle U_i^{lc}(w^*_{l,i})$, respectively. 
However, because of the local nature electrical synapses and non-locality of the chemical  synapses, we fixed $n_{_{l,e}}=1$ and $n_{_{l,c}}=8$ 
throughout our numerical computations. 
Hence, the two control parameters used are the synaptic time-delayed couplings ($\tau_{_{l,e}},\kappa_{_{l,e}}$) and ($\tau_{_{l,c}},\kappa_{_{l,c}}$).
On the other boundary, $\sigma^{\text{max}^e}_l$ and  $\sigma^{\text{max}^c}_l$ do not depend on the coordinates of the stable homogeneous fixed point, 
but on the complicated functions  $F_e(\kappa_{_{l,e}},\tau_{_{l,e}},n_{_{l,e}})$ and $F_c(\kappa_{_{l,c}},\tau_{_{l,c}},n_{_{l,c}})$, completely defined in Eqs. \eqref{eq:5}, \eqref{eq:6} and \eqref{eq:7}.
Knowing the minimum and maximum range of the noise amplitude within which SISR occurs will be very useful in discussing the numerical results in the following sections. 

\section{SISR in isolated layers}\label{section5}
\subsection{SISR in isolated layers with electrical synapses only}
\begin{figure}[htp!]
\begin{center}
\includegraphics[width=1.0\textwidth]{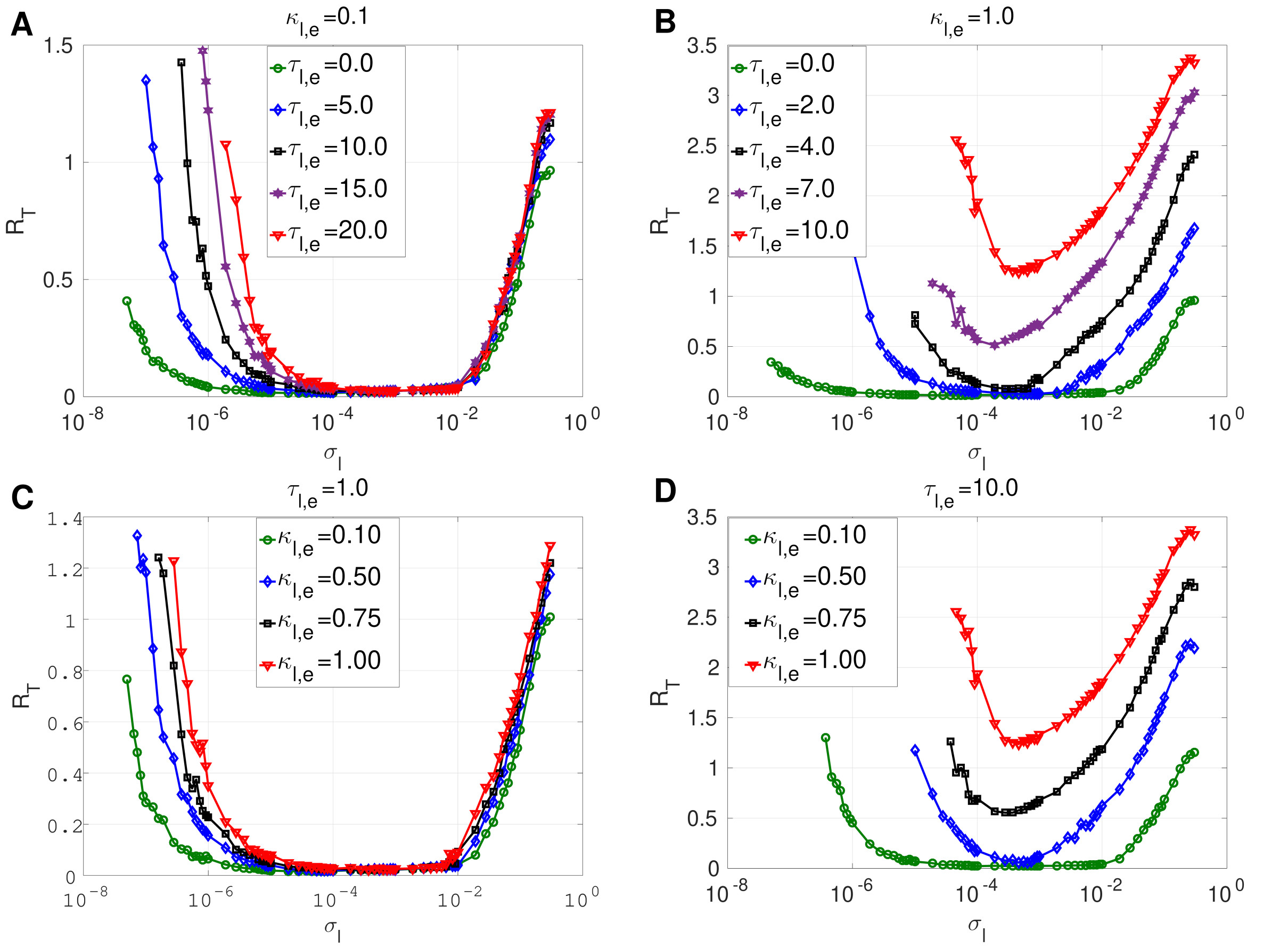}
\caption{Coefficient of variation $R_{_{T}}$ against noise amplitude $\sigma_{_l}$ of layer $l$ in isolation.
Increasing (decreasing) the electrical synaptic strength $\kappa_{_{l,e}}$ or the length of its time delay $\tau_{_{l,e}}$, 
deteriorates (enhances) SISR by increasing (decreasing) the values of $R_{_{T}}$ and by shrinking (extending) the interval of the noise amplitude in which $R_{_{T}}$ 
can achieve very low values. For example in panel \textbf{(D)}, for $\kappa_{_{l,e}}=1.0$ and $\tau_{_{l,e}}=10.0$, the red $R_{_{T}}$-curve lies 
entirely above the line $R_{_{T}}=1.0$ with a lowest value of $R_{_{T_\text{min}}}=1.24$ occurring at just one point $\sigma_{_l}=4.6\times10^{-4}$, indicating the non-existence of SISR.  
Parameters of layer $l$: $N=25$, $n_{_{l,e}}=1$, $\beta=0.75$, $\varepsilon=0.0005$, $\alpha=0.5$.\label{fig:4}}
\end{center} 
\end{figure}

We begin our numerical study with the dynamics of layer $l$ in isolation, where neurons are connected only via local electrical synapses in a ring network topology, i.e., we consider Eq.~\eqref{eq:1} 
with $n_{_{l,e}}=1$, $\kappa_{_{l,e}}\neq0$ and $\kappa_{_{m,e}}=\kappa_{_{m,c}}=\kappa_{_{l,c}}=0$. 
\textbf{Figure~\ref{fig:4}} shows the variation of $R_{_{T}}$ against the noise amplitude $\sigma_l$ for this layer. 
In the numerical computations, we choose $\varepsilon=0.0005\ll1$, because SISR can only occur in the singular limit, $\varepsilon\rightarrow0$, 
and the weak noise limit, $\sigma_l\rightarrow0$, imposed by Eq.~\eqref{eq:4}. 

In \textbf{Figure~\ref{fig:4}A}, a weak electrical synaptic strength is considered fixed, $\kappa_{_{l,e}}=0.1$. 
All the flat-bottom $R_{_{T}}$-curves obtained with different time delays 
($\tau_{_{l,e}}=0.0, \tau_{_{l,e}}=5.0, \tau_{_{l,e}}=10.0$, $\tau_{_{l,e}}=15.0$, $\tau_{_{l,e}}=20.0$) show a deep and 
broad minimum, indicating that the spike train has a high degree of coherence due to SISR for a wide range of the noise amplitude. 
We notice that even though the minimum (and low) values of $R_{_{T}}$ 
stays constant for various time delays, 
the left branch of the $R_{_{T}}$-curve is significantly being shifted to the right as the time delay increases. This means that with weak electrical synapses, the coherence of the spiking activity 
due to SISR is not affected as the time delay becomes longer, but the coherence is achieved only at relatively larger noise amplitudes $\sigma_{_l}$. Thus, we can obtain the same degree of SISR with 
longer time delays provided that we increase the noise amplitude (within the interval given in Eq.~\eqref{eq:8}) as the time delay increases. In \textbf{Figure~\ref{fig:4}A}, we have approximately the same minimum
value of $R_{_{T_\text{min}}}\approx0.015$ for: $\tau_{_{l,e}}=0.0$ with $\sigma_{_l}\in\big(3.7\times10^{-7},1.9\times10^{-2}\big)$; $\tau_{_{l,e}}=5.0$ with
$\sigma_{_l}\in\big(2.8\times10^{-6},1.9\times10^{-2}\big)$;
$\tau_{_{l,e}}=10.0$ with $\sigma_{_l}\in\big(5.5\times10^{-6},1.0\times10^{-2}\big)$;
$\tau_{_{l,e}}=15.0$ with $\sigma_{_l}\in\big(1.9\times10^{-5},1.0\times10^{-2}\big)$; 
and $\tau_{_{l,e}}=20.0$ with $\sigma_{_l}\in\big(2.8\times10^{-5},1.0\times10^{-2}\big)$. 
We note that the lower bound of the noise intervals increases as the time delay increases while the upper bounds are almost fixed.

In \textbf{Figure~\ref{fig:4}B}, we consider a strong electrical synapse ($\kappa_{_{l,e}}=1.0$). 
We observe that in contrast to \textbf{Figure~\ref{fig:4}A} with a weak electrical synapse, increasing the 
time delay squeezes the left and right branches of 
the $R_{_{T}}$-curves into a smaller noise interval, while shifting the curves to higher values, thus deteriorating SISR. 
In \textbf{Figure~\ref{fig:4}B}, 
we have different noise intervals for different minima of $R_{_{T}}$:
$R_{_{T_\text{min}}}=0.015$ at $\tau_{_{l,e}}=0.0$ for $\sigma_{_l}\in\big(2.8\times10^{-7},2.9\times10^{-2}\big)$;  
$R_{_{T_\text{min}}}=0.029$ at $\tau_{_{l,e}}=2.0$ for $\sigma_{_l}\in\big(2.8\times10^{-5},2.8\times10^{-3}\big)$; 
$R_{_{T_\text{min}}}=0.078$ at $\tau_{_{l,e}}=4.0$ for $\sigma_{_l}\in\big(1.9\times10^{-4},6.4\times10^{-4}\big)$.
In the last two cases, the noise intervals, in which we have the most deteriorated SISR, have shrunk to points with 
$R_{_{T_\text{min}}}=0.51$ at $\sigma_{_l}=1.9\times10^{-4}$ for $\tau_{_{l,e}}=7.0$; 
and $R_{_{T_\text{min}}}=1.24$ at $\sigma_{_l}=4.6\times10^{-4}$ for  $\tau_{_{l,e}}=10.0$. 
Thus, we see that with strong electrical synapses, the effect of the time delay on SISR becomes significant, unlike
when the electrical synapse is weak as in \textbf{Figure~\ref{fig:4}A}. 
In \textbf{Figure~\ref{fig:4}B}, we observe that even though the 
$R_{_{T}}$-curves for $\tau_{_{l,e}}=7.0$ and $\tau_{_{l,e}}=10.0$ are non-monotonic (characteristic of the existence of an optimal noise value for coherence), 
the minimum values of these curves are high ($0.51$ and $1.24$, respectively). 
Here, at only $\tau_{_{l,e}}=10.0$, $R_{_{T_\text{min}}}$ is already above 1.0 (indicating a stochastic spiking activity that is more variable than the Poisson 
process), whereas with weak electrical 
synapses in \textbf{Figure~\ref{fig:4}A}, even at $\tau_{_{l,e}}=20.0$, we still have $R_{_{T_\text{min}}}\approx0.015$.

In \textbf{Figure~\ref{fig:4}C} and \textbf{D}, we vary the electrical synaptic strength while the synaptic time is fixed  at a 
short ($\tau_{_{l,e}}=1.0$) and a long ($\tau_{_{l,e}}=10.0$) delay, respectively. A similar behavior as in \textbf{Figure~\ref{fig:4}A} 
and \textbf{B} is observed, with weak and strong electrical synaptic strengths, respectively. That is, at short synaptic time delays 
(see \textbf{Figure~\ref{fig:4}C}), the $R_{_{T}}$-curves show a deep and broad minimum, indicating a high degree of coherence due to SISR for a wide range of 
the noise amplitude when the  electrical synaptic strength $\kappa_{_{l,e}}$ is varied. 
Here, as $\kappa_{_{l,e}}$ increases, only the left branches of the $R_{_{T}}$-curves are shifted to the right, while the right branch of the $R_{_{T}}$-curves 
are fixed --- thereby fixing the upper bound of the noise amplitude $\sigma_{_l}$ below which SISR is optimal. 
This means that at short electrical time delays, the coherence of the spiking activity due to SISR is not affected as the electrical synaptic strength becomes 
stronger, but the coherence is achieved only at relatively larger noise amplitudes $\sigma_{_l}$. In \textbf{Figure~\ref{fig:4}D}, where electrical synaptic time 
delays are longer, increasing the electrical synaptic strength not only increases the minimum value of the 
$R_{_{T}}$-curves (thereby deteriorating SISR), but also shrinks the size of the noise interval in which SISR is optimized on both ends.

The response of SISR to changes in the synaptic strength $\kappa_{_{l,e}}$ and time delay $\tau_{_{l,e}}$ in \textbf{Figure~\ref{fig:4}} can be 
explained in terms of the electrical interaction potential $U^e_i(v_{_{l,i}},w_{_{l,i}})$ given in Eq.~\eqref{eq:4a} and represented in Fig.~\ref{fig2}. 
We observe in Fig.~\ref{fig2} that for a fixed ($n_{_{l,e}}=1$) ring network topology and time delay $\tau_{_{l,e}}$, 
as the synaptic strength $\kappa_{_{l,e}}$ increases, the energy barriers $\triangle U_i^{le}(w_{_{l,i}})$ and $\triangle U_i^{re}(w_{_{l,i}})$ 
become deeper. In particular, when $w_{_{l,i}}<0$, the trajectory is in the left potential well and as $\kappa_{_{l,e}}$ becomes stronger ($0.25, 0.5, 0.75, 1.0$), 
the left energy barrier $\triangle U_i^{le}(w_{_{l,i}})$ becomes deeper (hence the trajectory at the bottom of the well get closer to the homogeneous 
stable fixed point at $w^*_{_{l,i}}=-0.666651$). Thus, the deeper the left energy barrier $\triangle U_i^{le}(w_{_{l,i}})$ is (in other words, the stronger the 
electrical synaptic strength $\kappa_{_{l,e}}$ is), the closer is the trajectory to the stable fixed point and the further away is the neural system from the 
oscillatory regime. For the trajectory to jump over a high energy barrier $\triangle U_i^{le}(w_{_{l,i}})$, a stronger noise amplitude $\sigma_{_l}$ is of course needed. 
This is why in \textbf{Figure~\ref{fig:4}} as $\kappa_{_{l,e}}$ increases, the left branch of the $R_{_{T}}$-curve is shifted to the right, meaning that stronger noise amplitudes 
are required to induce frequent spiking (i.e., frequent escaping from the deep left energy barrier). But as the noise amplitude becomes bigger, the condition in 
Eq.~\eqref{eq:4} requiring $\sigma_{_l}\rightarrow0$ for the occurrence of SISR is violated. Hence, SISR disappears with increasing synaptic strength.

We can also see from \textbf{Figure~\ref{fig:4}D} that at longer time delay $\tau_{_{l,e}}$, this effect
(the shifting of the left branch of the $R_{_{T}}$-curve to the right) is more pronounced than 
in \textbf{Figure~\ref{fig:4}C} with a shorter time delay. 
This is because in Eq.~\eqref{eq:4a}, the longer the time delay is ($\tau_{_{l,e}}\gg0$), 
the further away is the quantity $\big[v_{_{l,i}}(t)v_{_{l,j}}(t-\tau_{_{l,e}})-v_{_{l,i}}(t)^2\big]$ from zero 
(since neurons are identical); hence, the stronger is the effect of the synaptic strength
$\kappa_{_{l,e}}$ on the electrical interaction potential, the energy barrier functions, 
and, consequently, on the $R_{_{T}}$-curves. 
Otherwise, if $\tau_{_{l,e}}\rightarrow0$, then because the neurons are identical,
$\big[v_{_{l,i}}(t)v_{_{l,j}}(t-\tau_{_{l,e}})-v_{_{l,i}}(t)^2\big]\rightarrow0$, 
and $\kappa_{_{l,e}}$ will have little effect on the electrical interaction potential, 
the energy barriers functions, and consequently on the $R_{_{T}}$-curves.
This is why the synaptic strength $\kappa_{_{l,e}}$ has a stronger effect on SISR 
only when $\tau_{_{l,e}}$ gets longer, and vice versa. This theoretical explanation will 
also support the behavior of the time-delayed chemical synapses in the optimization of SISR as we shall see further below.

Secondly, at weak electrical synaptic strengths and short time delays
(\textbf{Figure~\ref{fig:4}A} and \textbf{C}), 
the upper bound of the noise interval for which the $R_{_{T}}$-curves achieve their minima is almost constant. 
Here, only the lower bound of the noise intervals is shifted to the right. 
Whereas at strong electrical synaptic strengths and long time delays (\textbf{Figure~\ref{fig:4}B} and \textbf{D}), 
both the lower and upper bounds of the noise intervals are shifted to the right and to the left 
as $\tau_{_{l,e}}$  and $\kappa_{_{l,e}}$ increase, respectively. 
This has the overall effect of shrinking the noise interval in which the $R_{_{T}}$-curves achieve their minima to 
a single value of $\sigma_{_l}$. This behavior can be explained in terms of the 
minimum and maximum noise amplitudes between which SISR occurs obtained in Eq.~\eqref{eq:8}.
 
We observe from Eq.~\eqref{eq:8} that $\sigma^{\text{min}^e}_l$ depends on the fixed point coordinate 
$w^*_{_{l,i}}(\kappa_{_{l,e}},\tau_{_{l,e}},n_{_{l,e}})$ which in turn also depends on $\kappa_{_{l,e}},\tau_{_{l,e}},$ and $n_{_{l,e}}=1$. 
Therefore, changing $\kappa_{_{l,e}}$ and $\tau_{_{l,e}}$ will change the value of 
$w^*_{_{l,i}}(\kappa_{_{l,e}},\tau_{_{l,e}},n_{_{l,e}})$ which will in turn change the value of $\sigma^{\text{min}^e}_l$ 
via the energy barrier function $\triangle U_i^{le}(w^*_{_{l,i}})$.
Numerical computations indicate that $\sigma^{\text{min}^e}_l$ increases as $\kappa_{_{l,e}}$ and $\tau_{_{l,e}}$ 
increase (see \textbf{Figure~\ref{fig:4}}). 
On the other boundary, $\sigma^{\text{max}^e}_l$ does not depend on the coordinates of the homogeneous stable fixed point, 
but on the complicated function $F_e(\kappa_{_{l,e}},\tau_{_{l,e}},n_{_{2,e}})$,  fully determined by Eq.~\eqref{eq:5} and Eq.~\eqref{eq:6}. 
In \textbf{Figure~\ref{fig:4}A} and \textbf{C} (i.e., in the regimes of weak electrical synaptic strength and short time delays, respectively), 
we notice that $\sigma^{\text{max}^e}_l \approx 10^{-2}$ is nearly constant for all values of the time delay and electrical synaptic strength 
used. In~\cite{yamakou2018coherent}, where a single isolate FHN neuron is considered,
such fixation of the upper bound of the noise interval in which SISR occurs 
was already observed. In the case of a single isolated FHN neuron, the function $F_e$ in Eq.~\eqref{eq:5} 
takes a simple constant value $F_e=\dfrac{3}{4}$. 
This implies (for a fixed $\varepsilon=0.0005$) a 
fixed value for $\sigma^{\text{max}^e}_l=\big(3/2\cdot\log_e(\varepsilon^{-1})\big)^{1/2}$. 

In the case where a network of coupled FHN neurons is considered, 
the fixation of the upper bound of the noise interval 
for which SISR occurs can only be observed
if $F_e(\kappa_{_{l,e}},\tau_{_{l,e}},n_{_{2,e}})\rightarrow C$, where $C$ is a constant. 
In particular,  in a weak electrical synaptic regime ($\kappa_{_{l,e}}\rightarrow0$) and 
short time delay ($\tau_{_{l,e}}\rightarrow0$) regime (or more precisely, 
$[v_{_{l,i}}v_{_{l,j}}(t-\tau_{_{l,e}})-v_{_{l,i}}^2(t)]\rightarrow0$ as $\tau_{_{l,e}}\rightarrow0$, 
because all the neurons are identical), $F(\kappa_{_{l,e}},\tau_{_{l,e}},n_{_{l,e}})\rightarrow\dfrac{3}{4}$. 
In these regimes (see \textbf{Figure~\ref{fig:4}A} and \textbf{C}), we observe that $\sigma_{_l}^{\text{max}^e}\approx10^{-2}$, corresponding to the value 
obtained in~\cite{yamakou2018coherent} for the case of a single isolated FHN neuron ($\kappa_{_{l,e}}=0$).
In the regimes of strong coupling ($\kappa_{_{l,e}}\gg0$) and of long time delays ($\tau_{_{l,e}}\gg0\Rightarrow [v_{_{l,i}}v_{_{l,j}}(t-\tau_{_{l,e}})-v_{_{l,i}}^2(t)]\neq0$) 
shown in \textbf{Figure~\ref{fig:4}B} and \textbf{D}, the function $F_e$ in Eq.~\eqref{eq:5} 
is now strongly modified by the large values of $\kappa_{_{l,e}}$ and $\tau_{_{l,e}}$.
This is why in these regimes, the upper bound $\sigma_{_l}^{\text{max}^e}$ of the noise interval, for which SISR occurs,
is not any longer fixed, but shifted to the left as  $\tau_{_{l,e}}$ and $\kappa_{_{l,e}}$ take on larger values.
In the case of chemical synapses, as we shall see later, 
the same theoretical explanation holds for the shrinking, 
on both ends, of the interval of the noise amplitude in which SISR is optimized.
Later, we shall focus on layer $l=2$ with a non-existent SISR when it is in isolation ($\kappa_{_{2,e}}=1.0$ 
and $\tau_{_{2,e}}=10.0$; see the red curve in \textbf{Figure~\ref{fig:4}D} with $R_{_{T_\text{min}}}>1$) and then investigate which 
multiplexing configuration can best optimize SISR in this layer when it is multiplexed with layer $l=1$ when it already exhibits
pronounced SISR.

\subsection{SISR in isolated layers with inhibitory chemical synapses only}
\begin{figure}[htp!]
\begin{center}
\includegraphics[width=1.0\textwidth]{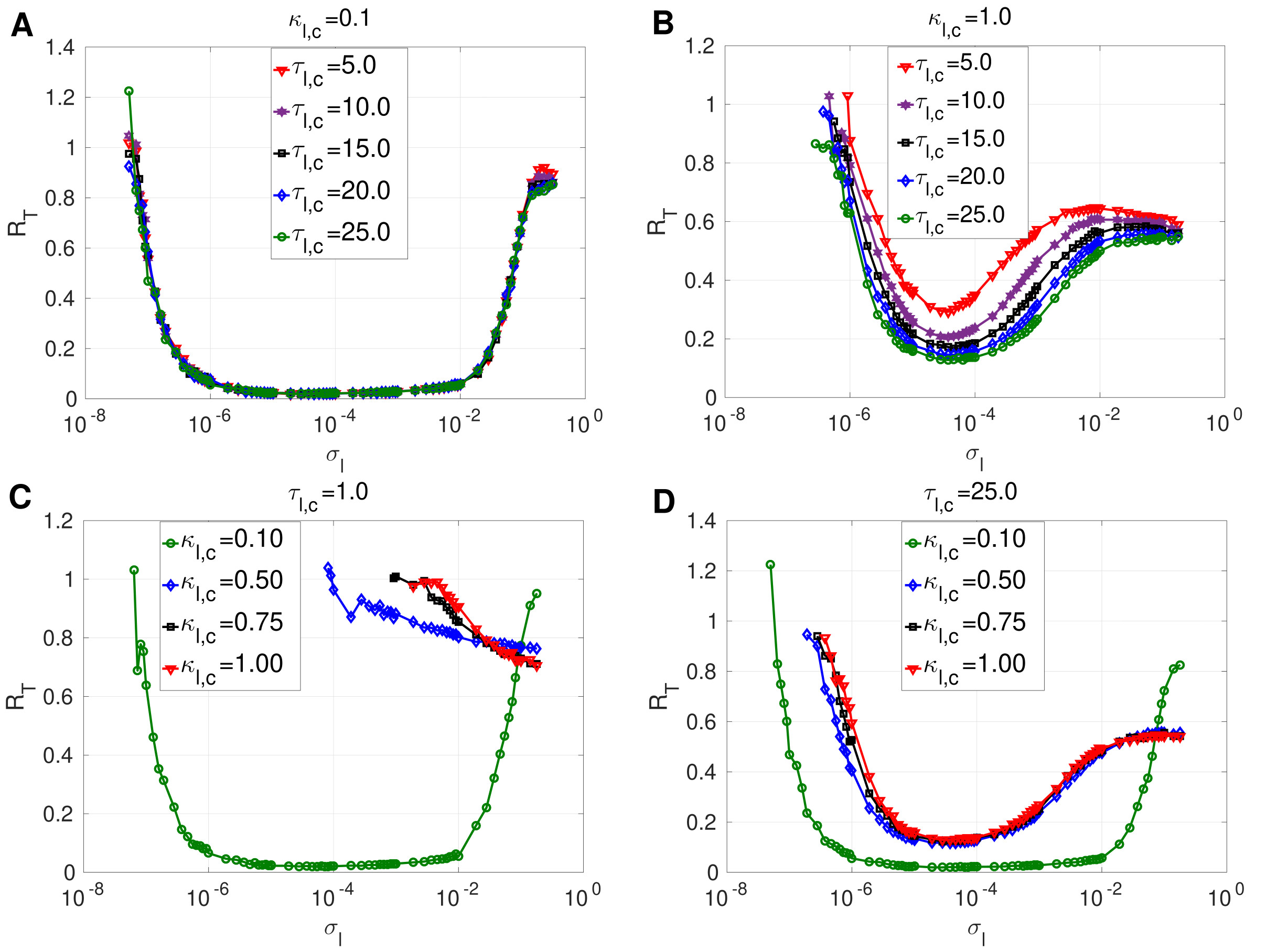}
\caption{Coefficient of variation $R_{_{T}}$ vs noise amplitude $\sigma_{_l}$ of layer $l$ in isolation.
Increasing (decreasing) the inhibitory chemical synaptic strength $\kappa_{_{l,c}}$ 
deteriorates (enhances) SISR by increasing (decreasing) the values of $R_{_{T}}$ and by shrinking (extending) 
the interval of the noise amplitude in which $R_{_{T}}$ can achieve very low values. Thus, inhibitory chemical  
synaptic strength qualitatively behaves as the electrical synaptic strength in optimizing SISR. 
However, electrical synaptic and inhibitory chemical synaptic time delays show opposite behaviors in the enhancement of SISR. 
Decreasing (increasing) the length of inhibitory chemical time delays $\tau_{_{l,c}}$, deteriorates (enhances) SISR by increasing 
(decreasing) the values of $R_{_{T}}$ and by shrinking (extending) the interval of the noise amplitude in which $R_{_{T}}$ can 
achieve very low values. This effect is particularly pronounced when the chemical synaptic strength is strong. For example in panel \textbf{(C)}, 
for $\kappa_{_{l,c}}=1.0$ and $\tau_{_{l,e}}=1.0$,  the red $R_{_{T}}$-curve achieves relatively high minimum value $R_{_{T_\text{min}}}=0.71$ 
occurring at a relatively large noise amplitude $\sigma_{_l}=4.6\times10^{-4}$, indicating a very poor SISR.  Parameters for layer $l$ are:
$N=25$, $n_{_{l,c}}=8$, $\beta=0.75$, $\varepsilon=0.0005$, $\alpha=0.5$.
\label{fig:5}}
\end{center} 
\end{figure}

We investigated the dynamics of layer $l$ in isolation, where neurons are connected only via (non-local) 
inhibitory chemical synapses in a ring network topology. Specifically, we consider Eq.~\eqref{eq:1} 
with $n_{_{l,e}}=8$, $\kappa_{_{l,c}}\neq0$ and $\kappa_{_{m,e}}=\kappa_{_{m,c}}=\kappa_{_{l,e}}=0$. \textbf{Figure~\ref{fig:5}} shows the variation 
of $R_{_{T}}$ against the noise amplitude $\sigma_l$ for this layer. 
We also fixed $\varepsilon=0.0005\ll1$ so that Eq.~\eqref{eq:4} can be satisfied in a weak noise limit $\sigma_l\rightarrow0$, leading 
to the occurrence of SISR. 
We shall now mainly compare the enhancement of SISR in layer $l$ for two situations, i.e., when the neurons are locally connected via time-delayed electrical 
synapses (see \textbf{Figure~\ref{fig:4}}) and when the neurons are non-locally connected via time-delayed inhibitory chemical 
synapses (see \textbf{Figure~\ref{fig:5}}).

The first observation is that longer inhibitory time delays enhance SISR, while longer electrical time delays deteriorate SISR. However, similarly to electric time delays, 
chemical time delays ($\tau_{_{l,c}}$) have a strong effect on SISR only for stronger chemical synaptic strength ($\kappa_{_{l,c}}$). 
In \textbf{Figure~\ref{fig:4}A}, the electrical synaptic strength is weak ($\kappa_{_{l,e}}=0.1$).
Even though the interval of the noise amplitude, for which a pronounced SISR occurs (as indicated by the very low values 
of $R_{_{T}}$), shrinks on the left bound with increasing time delay, the low values of $R_{_{T}}$ within that interval remain unchanged ($\approx0.015$). 
Similarly, results in \textbf{Figure~\ref{fig:5}A} with the same weak inhibitory synaptic strength ($\kappa_{_{l,c}}=0.1$),
show that changing the chemical time delays does also not affect the low and constant values of $R_{_{T}}\approx0.014$ (indicating an optimized SISR). In contrast, however, 
the lower bound of the noise interval with optimal SISR remains independent of varying levels of time delay. 
Thus, for weak synaptic strength and for increasing synaptic time delays, 
inhibitory chemical synapses outperform electrical synapses in optimizing SISR, in the sense that the former allow for a wider range of noise 
amplitudes for which $R_{_{T}}$ remains low. 

In \textbf{Figure~\ref{fig:5}B} where a large inhibitory chemical synaptic strength ($\kappa_{_{l,c}}=1.0$) is considered, 
time delays can have significant effect on SISR, in contrast to \textbf{Figure~\ref{fig:5}A} where $\kappa_{_{l,c}}$ is weak. 
In \textbf{Figure~\ref{fig:5}B},  increasing the chemical time delay enhances SISR by lowering the minimum value of $R_{_{T}}$.
In comparison to  \textbf{Figure~\ref{fig:5}A}, the noise interval for which SISR is optimal has shrunk on both sides. The reason for this shrinking on both ends of the optimal 
noise interval is essentially the same as for the case where the strength of electrical synapses is varied, see  \textbf{Figure~\ref{fig:4}B} and \textbf{D}. 
In \textbf{Figure~\ref{fig:5}B} we have
$R_{_{T_\text{min}}}=0.29$ at $\sigma_{_l}=3.7\times10^{-5}$ for $\tau_{_{l,c}}=5.0$; 
$R_{_{T_\text{min}}}=0.21$ at $\sigma_{_l}=3.7\times10^{-5}$ for $\tau_{_{l,c}}=10.0$;
$R_{_{T_\text{min}}}=0.17$ at $\sigma_{_l}=4.6\times10^{-5}$ for $\tau_{_{l,c}}=15.0$;
$R_{_{T_\text{min}}}=0.15$ at $\sigma_{_l}=4.6\times10^{-5}$ for $\tau_{_{l,c}}=20.0$;
$R_{_{T_\text{min}}}=0.12$ at $\sigma_{_l}=3.7\times10^{-5}$ for $\tau_{_{l,c}}=25.0$. 
However, the  deteriorating effects of electrical time delays on SISR is more pronounced than those of the chemical time delays 
at the same synaptic strength ($\kappa_{_{l,e}}=\kappa_{_{l,c}}=1.0$); see \textbf{Figure~\ref{fig:4}B} and  \textbf{Figure~\ref{fig:5}B}. 
This confirms that chemical synapses are better at optimizing SISR than electrical synapses, not only because they allow for a wider range of noise 
amplitude in which optimal SISR may occur, but also for the occurrence of a more enhanced SISR, as indicated by the relatively lower values of $R_{_{T}}$ at long time delays. 

In \textbf{Figure~\ref{fig:5}C} and \textbf{D} we investigate the effects of chemical synaptic strength in a short and long time delay regime.
Irrespective of the time delay regime, the stronger the chemical synaptic strength is, the more deteriorated is SISR. The reason 
behind this behavior is the same as the one given for the case of electrical synapses. That is, 
as the chemical synaptic strength $\kappa_{_{l,c}}$ becomes larger, the energy barriers $\triangle U_i^{lc}(w_{_{l,i}})$ 
and $\triangle U_i^{rc}(w_{_{l,i}})$ become deeper (see Fig.~\ref{fig3}). However, now a stronger noise amplitude is required to 
jump over the deep energy barriers and induce spiking, and this strong noise amplitude destroys the coherence of the spiking (by 
violating the conditions in Eq.~\eqref{eq:4} requiring $\sigma_{_l}\rightarrow0$) and hence deteriorates SISR. 

The deterioration of SISR by stronger chemical synaptic strengths is also observed with stronger electrical synaptic strength. 
However, for short synaptic time delay regimes ($\tau_{_{l,e}}=1.0=\tau_{_{l,c}}$, see \textbf{Figure~\ref{fig:4}C} 
and \textbf{Figure~\ref{fig:5}C}), we notice the following difference for both synaptic types:
When the time delay is relatively short, electrical synapses optimize SISR compared to chemical synapses as the synaptic strength is weakened.
We see in \textbf{Figure~\ref{fig:5}C} with $\tau_{_{l,c}}=1.0$ that SISR is destroyed as the $\kappa_{_{l,c}}$ increases, 
whereas in \textbf{Figure~\ref{fig:4}C} with $\tau_{_{l,e}}=1.0$, SISR remains enhanced as $\kappa_{_{l,e}}$ increases.
This means that an electrical synapse is a better means than a chemical synapse in optimizing SISR at very short time delays, 
irrespective of the synaptic strengths, while a chemical synapse is better than a electrical synapse 
at very long time delays, irrespective of the synaptic strengths.

In \textbf{Figure~\ref{fig:5}}, the reason for the deterioration of SISR with decreasing time delays could be inferred from the reason given for 
the deterioration of SISR with increasing synaptic strength. That is, shortening the chemical synaptic time delays 
increases the depth of the chemical energy barrier functions given in Eq.~\eqref{eq:5}. This will in turn demand larger noise amplitude to jump over deep energies 
barriers to induce spiking with no coherence, and hence very poor SISR; as seen, for example, from the red curve in \textbf{Figure~\ref{fig:5}C}. 
Here, we see that rare spiking can be induced only when the noise $\sigma_{_l}\geq10^{-4}$ as $R_{{T}}$ stays high with increasing noise 
amplitude $\sigma_{_l}$.
However, from conditions in Eq.~\eqref{eq:4}, SISR requires  $\sigma_{_l}\rightarrow0$, which implies that increasing the noise would not improve SISR.
We can see from the red curve  in \textbf{Figure~\ref{fig:5}C} that a minimum value of $R_{_{T_\text{min}}}\approx0.71$, 
already indicating a very poor SISR, 
occurs at a relatively large noise amplitude of $\sigma_{_l}=0.18$.
Below, we shall focus on the enhancement of this very poor SISR in layer $l=2$ (for $\kappa_{_{2,c}}=\tau_{_{2,c}}=1.0$; see also red curve in 
\textbf{Figure~\ref{fig:5}C}) by using various 
multiplexing configurations, where layer $l=1$ already exhibits enhanced SISR in isolation.

\section{Multiplexing and optimization of SISR}\label{section6}
We now address the questions: 
(1) Is an optimization of SISR based on the multiplexing of layers 
possible? 
(2) Which synaptic multiplexing configuration is the best optimizer of SISR?
To answer these two questions, we configure the synaptic strength and time delay of one layer (say layer 1) such 
that SISR is optimal in this layer. The corresponding parameters of layer 2 are configured such that SISR is non-existent in this layer.
Then, we connect these two layers in a multiplex fashion (see \textbf{Figure~\ref{fig:1}}) in six different multiplexing configurations. 

In the first three configurations, the two layers of the multiplex network each consist of neurons that 
are intra-connected by only electrical synapses ($\kappa_{_{1,e}}, \tau_{_{1,e}}$) and ($\kappa_{_{2,e}}, \tau_{_{2,e}}$), 
and 
are inter-connected (multiplexed) by
(i) electrical synapses ($\kappa_{_{m,e}}, \tau_{_{m,e}}$), 
(ii) inhibitory chemical synapses ($\kappa_{_{m,c}}, \tau_{_{m,c}}$), and  
(iii) excitatory chemical synapses ($\kappa_{_{m,c}}, \tau_{_{m,c}}$). 

In the next three configurations, we use the same three synaptic multiplexing configurations of two layers, each consisting of neurons that are 
intra-connected by only inhibitory chemical synapses. We do not consider excitatory chemical synapses for intra-connectivity, because this type of synapse induces 
coherent spiking activities even in the absence of noise -- which is not a requirement for SISR. 
On the other hand, we use excitatory chemical synapses in the inter-layer connections (multiplexing),
because for some synaptic strengths and time delays the multiplex network remains in the excitable regime in the absence of noise -- 
a requirement for observing SISR. However, we also have situations in which the multiplexing excitatory chemical synapses
strengthen the excitable regime of the network by making the homogeneous fixed point more stable --- thereby requiring very large noise amplitudes to 
have a chance of inducing a spike. But large noise amplitudes violates the conditions necessary for the occurrence of SISR. 
So we shall also avoid such regimes and stay in the excitable regimes, where vanishingly small noise amplitudes have a non-zero probability of inducing 
at least a spike in the large time interval we considered in our simulations.

In the following numerical simulations, we have ensured that the multiplex network stays in the excitable regime 
by checking that all the synaptic strengths and time delays of the multiplexing synapses are such that no 
self-sustained spiking activity occurs in the absence of noise. It is worth mentioning 
here that the optimization of SISR based on the multiplexing approach appears not to be feasible in a network of mixed layer type, i.e., 
consisting of an electrical layer multiplexed to a chemical layer. 
We investigated all the possible configurations of mixed 
layered networks, i.e., those with electrical, inhibitory, and excitatory chemical multiplexing. 
Extensive numerical simulations (not shown) clearly indicated that the optimization of SISR, 
for the ranges of the multiplexing synaptic strengths and time delays considered, was not possible in mixed layered networks. 
For this reason, we only discuss the layered networks that display the capability of optimizing SISR and compare the optimization abilities of various multiplexing connections between two electrical layers and then between two inhibitory chemical layers.

\subsection{Multiplexing of electrical layers}
\begin{figure}[htp!]
\begin{center}
\includegraphics[width=1.0\textwidth]{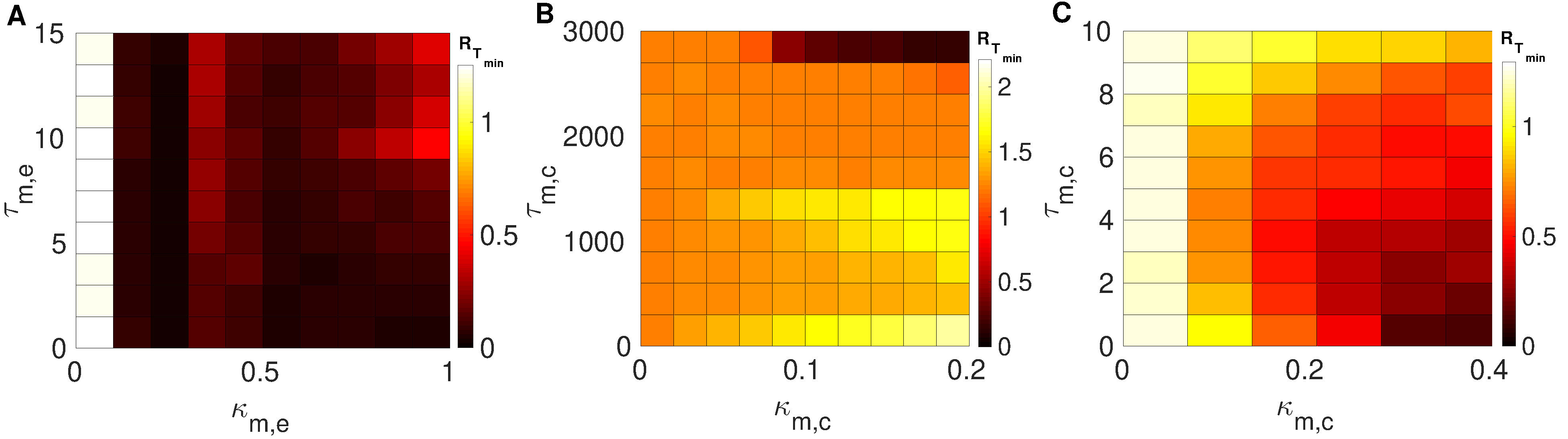}
\caption{Color-coded minimum coefficient of variation ($R_{_{T_\text{min}}}$) of layer 2 as a function of the multiplexing parameters.
Both layers 1 and 2 are intra-connected by electrical synapses. Panel \textbf{(A)} shows the enhancement performances of 
the electrical multiplexing $(\kappa_{_{m,e}},\tau_{_{m,e}})$.
For optimized SISR in layer 2, we need either a shorter $\tau_{_{m,e}}$ and stronger $\kappa_{_{m,e}}$; or a longer $\tau_{_{m,e}}$ and a weaker $\kappa_{_{m,e}}$. 
Panel \textbf{(B)} shows the enhancement performances of inhibitory chemical multiplexing $(\kappa_{_{m,c}},\tau_{_{m,c}})$. 
For optimized SISR in layer 2, we need a stronger $\kappa_{_{m,c}}$ and a very long $\tau_{_{m,c}}$.
Panel \textbf{(C)} shows the enhancement performances of excitatory chemical multiplexing $(\kappa_{_{m,c}},\tau_{_{m,c}})$. 
For optimized SISR in layer 2 we need: very short $\tau_{_{m,c}}$ and stronger $\kappa_{_{m,c}}$. 
Parameters for layer 1: $N=25$, $n_{_{1,e}}=1$, $\beta=0.75$, $\varepsilon=0.0005$, $\alpha=0.5$, $\kappa_{_{1,e}}=0.1$,  $\tau_{_{1,e}}=1.0$.
Parameters for layer 2: $N=25$, $n_{_{2,e}}=1$, $\beta=0.75$, $\varepsilon=0.0005$, $\alpha=0.5$, $\kappa_{_{2,e}}=1.0$,  $\tau_{_{2,e}}=10.0$.
\label{fig:6}}
\end{center} 
\end{figure}

We consider layer 1 and layer 2 in which neurons are electrically coupled only.
We choose the synaptic parameters ($\kappa_{_{1,e}}, \tau_{_{1,e}}$) of layer 1 such that SISR is pronounced, i.e., 
we choose a weak synaptic strength $\kappa_{_{1,e}}=0.1$ and a short synaptic time 
delay $\tau_{_{1,e}}=1.0$ (green $R_{_{T}}$-curve \textbf{Figure~\ref{fig:4}C} with  $R_{_{T_\text{min}}}=0.015$). 
For layer 2, we set the synaptic parameters such that SISR is non-existent, 
i.e., we choose a strong synaptic strength  $\kappa_{_{1,e}}=1.0$ and a long synaptic time delay $\tau_{_{2,e}}=10.0$ 
(red $R_{_{T}}$-curve \textbf{Figure~\ref{fig:4}D} with $R_{_{T_\text{min}}}=1.24$). These two layers are then coupled in a 
multiplex network as shown in \textbf{Figure~\ref{fig:1}A}.  The multiplexing introduces two other parameters -- the multiplexing synaptic strengths 
$\{\kappa_{_{m,e}},\kappa_{_{m,c}}\}$ and their corresponding time delays $\{\tau_{_{m,e}},\tau_{_{m,c}}\}$. \textbf{Figure~\ref{fig:6}} 
shows the color-coded minimum values of the coefficient of variation $R_{_{T_\text{min}}}$ of layer 2 as a function of the multiplexing parameters for 
the three multiplexing configurations considered.

In \textbf{Figure~\ref{fig:6}A}, the multiplexing between the two layers is mediated by electrical synapses with 
parameters $(\kappa_{_{m,e}},\tau_{_{m,e}})$. We can clearly see that even very weak multiplexing $\kappa_{_{m,e}}\geq0.1$, particularly at 
short time delays $\tau_{_{m,e}}\leq9.5$, can induce a very pronounced SISR in layer 2 (where SISR was non-existent in isolation) as indicated by the 
dark red color corresponding to very low values of $R_{_{T_\text{min}}}$. In the region $\tau_{_{m,e}}\leq9.5$, stronger multiplexing strengths push $R_{_{T_\text{min}}}$ 
to even lower values as indicated by the darker red color, thus optimizing SISR in layer 2.
However, as the multiplexing time delay becomes longer $\tau_{_{m,e}}>9.5$, this time delay starts to dominate the control of SISR. 
As the time delay $\tau_{_{m,e}}>9.5$ increases, 
SISR progressively deteriorates and the effect of strong multiplexing is reversed, i.e., the stronger $\kappa_{_{m,e}}$ is, 
the more SISR deteriorates, as indicated by the change of color of $R_{_{T_\text{min}}}$ from dark red to light red. 
While in this same region, i.e.,  $\tau_{_{m,e}}>9.5$, weaker multiplexing optimize SISR better than strong ones,
as seen in the region bounded by $\tau_{_{m,e}}\in[9.5,15.0]$ and $\kappa_{_{m,e}}\in[0.1,1.0]$ with a dark red color.

In \textbf{Figure~\ref{fig:6}B}, the multiplexing between the two electrical
layers is mediated by inhibitory chemical synapses with parameters $(\kappa_{_{m,c}},\tau_{_{m,c}})$. We notice, in contrast to \textbf{Figure~\ref{fig:6}A},
that the multiplexing inhibitory chemical synaptic strength takes a maximum value of 
$\kappa_{_{m,c}}=0.2$, i.e., it stays in the weak multiplexing regime and the time delay goes up to the very large value of $\tau_{_{m,c}}=3000$.
As already pointed out, we always want the network to stay in the excitable regime such that self-sustained oscillations do not arise due to bifurcations.
In this multiplexing configuration, values of the inhibitory chemical synaptic strength greater than $0.2$ induces oscillations in the absence of noise ---
for SISR, the system should oscillate coherently due only to the presence of noise and not due to a bifurcation. As we can see in \textbf{Figure~\ref{fig:6}B}, 
weak multiplexing inhibitory chemical synaptic strength $\kappa_{_{m,c}}\in[0.0,0.2]$ cannot induce SISR in layer 2 as the values of $R_{_{T_\text{min}}}$  stay very high above 1.0, 
except at very long multiplexing delays $\tau_{_{m,c}}\geq2750$. 
It can be observed that for time delays $\tau_{_{m,c}}\leq1500$, stronger multiplexing values, $\kappa_{_{2,c}}\gtrsim0.1$, deteriorate SISR (yellow region) to a 
larger extent than the weaker values, $\kappa_{_{2,e}}\lesssim0.1$ (orange region). 
But when the multiplexing time delay become very long, e.g., at $\tau_{_{m,c}}=3000$, 
stronger multiplexing ($\kappa_{_{2,c}}\gtrsim0.1$) induces an optimized (dark red color of $R_{_{T_\text{min}}}$) 
SISR in layer 2, while weaker multiplexing ($\kappa_{_{2,c}}\lesssim0.1$) cannot optimize SISR, as indicated by the orange color of $R_{_{T_\text{min}}}$.
This means that multiplexing with inhibitory chemical synapses has the opposite effect compared to multiplexing with electrical synapses, in terms  of SISR in layer 2.
To sum up, stronger $\kappa_{_{m,c}}$ means poorer SISR at shorter $\tau_{_{m,c}}$, but better SISR at longer $\tau_{_{m,c}}$; 
vice versa, stronger $\kappa_{_{m,e}}$ means better SISR at shorter $\tau_{_{m,e}}$, but poorer SISR at longer $\tau_{_{m,e}}$.

In \textbf{Figure~\ref{fig:6}C}, the multiplexing between the two electrical layers is mediated by excitatory chemical synapses with 
parameters $(\kappa_{_{m,c}},\tau_{_{m,c}})$. First, we notice the range of the synaptic strength and the time delay. For $\kappa_{_{m,c}}>0.4$, 
the excitability of the network becomes so strong that even large noise amplitudes (SISR requires vanishingly small noise) are not longer 
capable of inducing a spike in the large time interval considered. In contrast to weak multiplexing electrical synapses in \textbf{Figure~\ref{fig:6}A},
weak multiplexing excitatory chemical synapses are incapable of inducing SISR in layer 2. 
In \textbf{Figure~\ref{fig:6}C}, for a weak multiplexing 
$\kappa_{_{m,c}}\in[0.0,0.28]$, $R_{_{T_\text{min}}}$ remains high with the 
lowest value above $0.5$ (as indicated by the white, yellow, orange and light red colors of $R_{_{T_\text{min}}}$) for all of 
the time delays considered. 
This inability of weak excitatory chemical multiplexing to optimize 
SISR in layer 2 is similar to that of weak inhibitory chemical multiplexing in \textbf{Figure~\ref{fig:6}B}. 
However, for an intermediate excitatory chemical multiplexing, i.e., $\kappa_{_{m,c}}\in]0.28,0.4]$, an optimized SISR is induced in layer 2 
(just like with intermediate electrical multiplexing in \textbf{Figure~\ref{fig:6}A}), 
but only at very short time delays $\tau_{_{m,c}}\in[0.0,2.0]$, 
where $R_{_{T_\text{min}}}$ assumes low values corresponding to the dark red colors of $R_{_{T_\text{min}}}$. 
And the main difference between inhibitory chemical multiplexing and excitatory chemical multiplexing is in terms of their time delays. 
While inhibitory chemical  multiplexing requires extremely long time delay ($\tau_{_{m,c}}\geq2750$) to optimize SISR in layer 2, excitatory chemical
multiplexing requires extremely short time delays ($\tau_{_{m,c}}\in[0.0,2.0]$) for the optimization.  

\subsection{Multiplexing of inhibitory chemical layers}
\begin{figure}[htp!]
\begin{center}
\includegraphics[width=1.0\textwidth]{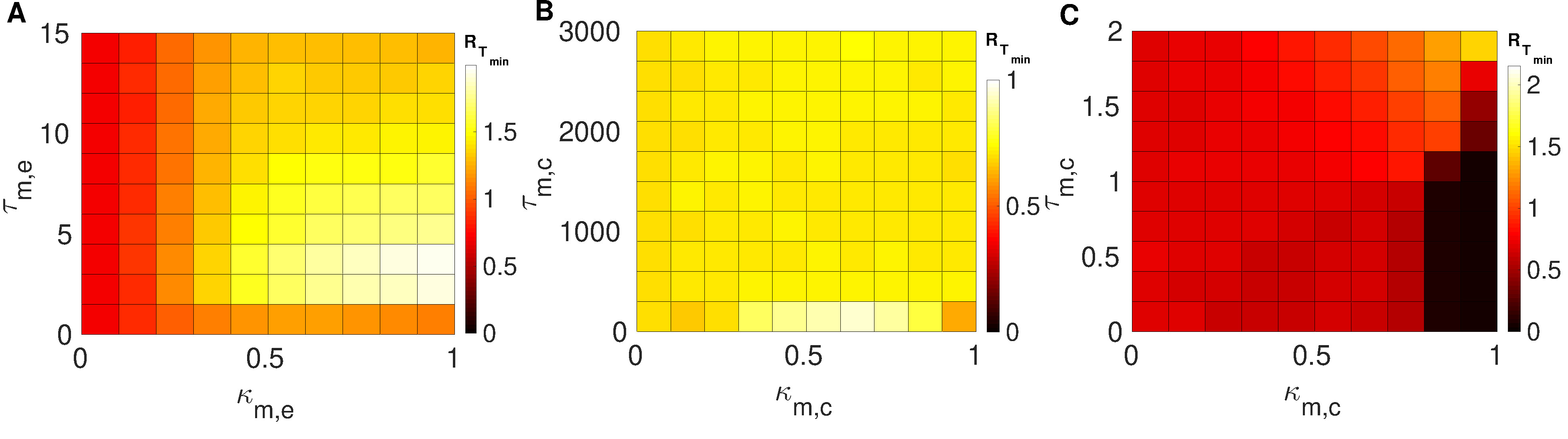}
\caption{Color-coded minimum coefficient of variation ($R_{_{T_\text{min}}}$) of layer 2 against multiplexing parameters.
Each of layer 1 and layer 2 is intra-connected by inhibitory chemical synapses.
Panel \textbf{(A)} shows the enhancement performances of the electrical multiplexing $(\kappa_{_{m,e}},\tau_{_{m,e}})$. 
Here, we observe that 
electrical multiplexing cannot optimize SISR in layer 2, especially at stronger multiplexing. 
Panel \textbf{(B)} shows the enhancement performances of the inhibitory chemical multiplexing $(\kappa_{_{m,c}},\tau_{_{m,c}})$. In this case,
the enhancement of SISR is even worse than in panel \textbf{(A)}, especially at intermediate multiplexing strengths and short time delays.
Panel \textbf{(C)} shows the enhancement performances of the excitatory chemical multiplexing $(\kappa_{_{m,c}},\tau_{_{m,c}})$. Here, an enhancement is possible.
An optimized SISR ($R_{_{T_\text{min}}}\approx 0.03$) emerging at strong excitatory chemical synapses ($\kappa_{_{m,c}}\geq0.8$) with short time delays ($\tau_{_{m,c}}\leq1.2$). 
Parameters of layer 1: $N=25$, $n_{_{1,c}}=8$, $\beta=0.75$, $\varepsilon=0.0005$, $\alpha=0.5$, $\kappa_{_{1c}}=0.1$,  $\tau_{_{1,c}}=25.0$.
Parameters of layer 2: $N=25$, $n_{_{l,c}}=8$, $\beta=0.75$, $\varepsilon=0.0005$, $\alpha=0.5$, $\kappa_{_{l,c}}=1.0$,  $\tau_{_{l,c}}=1.0$. 
\label{fig:7}}
 \end{center} 
 \end{figure}

Here we consider layers 1 and 2 in which neurons are coupled only via inhibitory chemical synapses. 
We set the synaptic parameters ($\kappa_{_{1,c}}, \tau_{_{1,c}}$) 
of layer 1 such that SISR is pronounced, i.e., we choose a weak synaptic strength 
$\kappa_{_{1,c}}=0.1$ and a long synaptic time delay $\tau_{_{1,c}}=25.0$, see the green $R_{_{T}}$-curve 
\textbf{Figure~\ref{fig:5}D} with  $R_{_{T_\text{min}}}=0.015$. 
For layer 2, we set the synaptic parameters such that SISR is very poor, i.e., we choose a strong synaptic strength 
$\kappa_{_{1,c}}=1.0$ and a short synaptic time delay 
$\tau_{_{2,c}}=1.0$, see the red $R_{_{T}}$-curve \textbf{Figure~\ref{fig:5}C} with $R_{_{T_\text{min}}}=0.71$. 
These two layers are then coupled in a multiplex network as shown in \textbf{Figure~\ref{fig:1}B}.
The multiplexing introduces two other parameters -- the multiplexing synaptic strengths 
$\{\kappa_{_{m,e}},\kappa_{_{m,c}}\}$ and their corresponding time delays $\{\tau_{_{m,e}},\tau_{_{m,c}}\}$. 
\textbf{Figure~\ref{fig:7}} shows the color-coded minimum values of the coefficient of variation $R_{_{T_\text{min}}}$ of layer 2 as a
function of the multiplexing parameters for the three multiplexing configurations considered.

In \textbf{Figure~\ref{fig:7}A}, the multiplexing between the two inhibitory chemical layers is mediated by 
electrical synapses with parameters $(\kappa_{_{m,e}},\tau_{_{m,e}})$. 
It is observed that electrical multiplexing cannot, at all optimize SISR in layer 2 as indicated by the 
very high values of $R_{_{T_\text{min}}}$ in the entire $\kappa_{_{m,e}}-\tau_{_{m,e}}$ parameter space. 
Comparing \textbf{Figure~\ref{fig:6}A} and \textbf{Figure~\ref{fig:7}A}, we can conclude that 
electrical multiplexing  become good optimizers of SISR only when the multiplexed layers are both
intra-connected by electrical synapses. In particular, we observe that, while a strong 
multiplexing $\kappa_{_{m,e}}$ with a short delay $\tau_{_{m,e}}$ of layers intra-connected by electrical synapses
optimizes SISR in layer 2, (see \textbf{Figure~\ref{fig:6}A}), a strong  multiplexing $\kappa_{_{m,e}}$ with a short delay $\tau_{_{m,e}}$ of 
layers intra-connected by inhibitory chemical synapses makes SISR rather worst ($R_{_{T_\text{min}}}\geq1.0$, 
see \textbf{Figure~\ref{fig:7}A}) in the layer 2 than when this layer is in isolation ($R_{_{T_\text{min}}}=0.71$).
 
In \textbf{Figure~\ref{fig:7}B}, the multiplexing between the two inhibitory chemical
layers is mediated by inhibitory chemical synapses with parameters $(\kappa_{_{m,c}},\tau_{_{m,c}})$. In this
multiplexing configuration, an optimization of SISR in layer 2 is impossible as well, especially at intermediate 
multiplexing strengths and short time delays, where the $R_{_{T_\text{min}}}$ assumes
an even larger value ($R_{_{T_\text{min}}}\approx0.9$) than layer 2 in isolation ($R_{_{T_\text{min}}}=0.71$). 
Moreover, even very long multiplexing time delays, as in the case of electrical layers multiplexed by inhibitory chemical 
synapses (see \textbf{Figure~\ref{fig:6}B}), 
cannot optimize SISR in layer 2,  irrespective of the multiplexing strength. 
Thus, we conclude that multiplexing inhibitory chemical synapses is generally a bad optimizer
of SISR in layers intra-connected by either chemical synapses or electrical synapses. 
However, recall that inhibitory chemical synapses can be very good 
optimizers of SISR within a layer, see \textbf{Figure~\ref{fig:5}}.

In \textbf{Figure~\ref{fig:7}C}, the multiplexing between the two inhibitory chemical
layers is mediated by excitatory chemical synapses with parameters $(\kappa_{_{m,c}},\tau_{_{m,c}})$. Note 
that the range of multiplexing time delay we considered is very short, i.e., $\tau_{_{m,c}}\in[0.0,2.0]$. 
For $\tau_{_{m,c}}>2.0$, excitatory chemical multiplexing induces self-sustained oscillations in the absence of noise --- a regime not required 
for SISR. In contrast to electrical and inhibitory chemical multiplexing of layers intra-connected with chemical synapses 
(see \textbf{Figure~\ref{fig:7}}\textbf{A} and \textbf{B}), excitatory chemical multiplexing of such layers can perform extremely well at optimizing SISR. 
However, this capability for very strong optimization is only possible at strong excitatory chemical multiplexing ($\kappa_{_{m,c}}\geq0.8$) 
and very short time delays ($\tau_{_{2,c}}\in[0.0,1.2]$) with $R_{_{T_\text{min}}}\approx 0.03$. 
This implies that excitatory chemical synapses, as a multiplexing synapse, could play more 
important functional roles (than electrical and inhibitory chemical synapses) 
in neural information processing due to SISR in multiplexed layers intra-connected by 
inhibitory chemical synapses.

\section{Conclusion}\label{section7}
We have investigated the effects of electrical and chemical synaptic couplings on the noise-induced phenomenon of SISR 
in isolated layers as well as in multiplexed layer networks of  
the FHN neuron model in the excitable regime. We have presented the analytic conditions necessary for SISR to 
occur in isolated layers with neurons connected either via electrical or 
inhibitory chemical synapses. From these analytic conditions, we have also obtained the minimum and maximum synaptic 
noise amplitude required for the occurrence of SISR in isolated layers.

Numerical computations indicate that in an isolated layer, the weaker the electrical synaptic strength and the shorter the 
corresponding synaptic time delay are, the more enhanced SISR is. 
However, the deteriorating effect of stronger electrical synaptic couplings is significant only at longer time delays and vice versa. 
On the other hand, in an isolated layer with inhibitory chemical synapses, 
weaker inhibitory chemical synaptic couplings just like their weaker electrical counterparts enhance SISR. 
Moreover, the longer the synaptic time delay is, the more enhanced is SISR --- in contrast to isolated layers with electrical synapses. 
The enhancing effect of the longer synaptic time delays in isolated layers with inhibitory chemical synapses becomes significant only 
at stronger synaptic strengths. 
Furthermore, it is also found that at very short time delays and irrespective of the synaptic strengths, electrical
synapses are better optimizers of SISR than chemical synapses. While at very long time delays and irrespective of
the synaptic strengths, chemical synapses are a better optimizers of SISR than electrical synapses.
The expressions of electrical and chemical 
interaction potentials together with the minimum and maximum values of the noise amplitude within which an optimized SISR can occur 
are used to provide a theoretical explanation of the 
above effects.

After identifying the electrical and chemical synaptic strengths and time delays that destroy (or optimize) SISR in an isolated layer, 
we proceeded with identifying multiplexing 
configurations between the two layers that would optimize SISR in the second layer where SISR would be very poor or non-existent in isolation. 
For this identification, the synaptic parameters of one layer is configured such that SISR is optimal and this layer is multiplexed with a second layer where synaptic parameters are such that 
SISR is very poor or even non-existent. We then investigated which multiplexing 
connection (i,e., electrical, inhibitory chemical, or excitatory chemical synapses) is a better 
optimizer of SISR in the second layer.

In the first optimization configuration, we were interested in optimizing SISR in an electrically coupled layer 
(i.e., a layer where neurons are coupled only via electrical synapses)
by multiplexing this layer with another electrically coupled layer. We found that even weak multiplexing with electrical synaptic 
connections may optimize SISR in 
the layer where SISR was even absent in isolation. However, the longer the multiplexing electrical 
synaptic time delay is, the less efficient this configuration becomes 
in optimizing SISR. In a second scenario, the multiplexing connection was mediated by inhibitory chemical 
synapses between these electrical layers. Here, we found that only very 
long multiplexing inhibitory chemical synaptic time delays at weak (but not too weak) synaptic strength may optimize SISR in the layer 
where it was non-existent in isolation. 
And in the third scenario, the multiplexing connection was mediated by excitatory chemical synapses between 
these electrical layers. It is found that only very short multiplexing excitatory 
chemical time delays at intermediate synaptic strengths can optimize SISR in the layer where the phenomenon is non-existent in isolation.

In the second optimization configuration, we were interested in optimizing SISR in an (inhibitory) chemically coupled 
layer (i.e., a layer where neurons are coupled only via inhibitory chemical synapses)
by multiplexing this layer with another (inhibitory) chemically coupled layer. Here it is found that the
optimization of SISR based multiplexing between chemical layers does work equally well as in the case of the 
multiplexing between electrical layers. Multiplexing of the chemical layers by electrical synapses and inhibitory chemical synapses 
cannot optimize SISR at all in the chemical layer, where 
in isolation SIRS is otherwise very poor. 
We found that only multiplexing excitatory chemical synapses (using a strong synaptic coupling and short time delay regime) 
can optimize SISR in the chemical layer, where in isolation SISR is very poor.

Comparing the first and the second optimization configurations of SISR, 
we conclude that the optimization of SISR is generally better in layers with electrically coupled neurons rather 
than with chemically coupled neurons, provided that multiplexing connections between the layers  are either electrical 
or an inhibitory chemical synapses. Vice versa, optimization of SISR is generally better in layers with (inhibitory) 
chemically coupled neurons than with electrically coupled neurons, when multiplexing connections between the layers 
are excitatory chemical synapses.

The manipulation of chemical and electrical patterns in the brain has become more accessible, either 
via drugs that cross the blood brain barrier, via electrical 
stimulation delivered through electrodes implanted in the brain, or via light delivered through optical fibers 
selectively exciting genetically manipulated 
neurons~\cite{de2017multilayer,andreev2018coherence,runnova2016theoretical}; however, the manipulation of the 
functional connectivity seems to be a more difficult goal to achieve. 
Our approach of modeling multi-layer networks in combination with stochastic dynamics  offers a novel perspective 
on the modeling of the brain's structural and functional connectivity. 
We therefore expect that our findings could provide promising applications in controlling synaptic connections to 
optimize neural information (generated by noise-induced phenomena like SISR and CR) 
processing in experiments, surgery involving brain networks stimulation, 
and in designing networks of artificial neural circuits to optimize information processing via SISR 
\cite{moopenn1990digital,moopenn1989digital,eberhardt1989vlsi}.

Interesting future research directions on the topic would be to investigate the optimization performances of electrical 
and chemical synapses 
in other intra-layer topologies like small-world network, scale-free network and random network; and other inter-layer 
topologies like the multiplex topology in which 
neurons in one layer are connected (randomly) to more than one neuron in the other layer. 
 
\section*{Acknowledgments}
M.E.Y. would like to acknowledge the warm hospitality at Department of Applied Mathematics and Computer 
Science of the Technical University of Denmark and for financial support.

\bibliography{refs}{}
\bibliographystyle{unsrt}
\end{document}